\documentclass{emulateapj}

\usepackage{float}
\usepackage{amsmath}
\usepackage{subfigure}
\usepackage{epsfig,floatflt}

\shorttitle{Testing for non-Gaussianity in the \emph{WMAP} data}
\shortauthors{Eriksen et al.}

\received{2004 January 14}
\accepted{2004 May 10}
\begin{document}

\title{Testing for Non-Gaussianity in the \emph{Wilkinson Microwave
Anisotropy Probe} Data: Minkowski
Functionals and the Length of the Skeleton}

\author{H.\ K.\ Eriksen\altaffilmark{1,2,3},  D.\
I. Novikov\altaffilmark{4}, and P.\ B.\ Lilje\altaffilmark{1}}

\affil{Institute of Theoretical Astrophysics, University of Oslo,
P.O.\ Box 1029 Blindern, N-0315 Oslo, Norway}

\email{h.k.k.eriksen@astro.uio.no} 
\email{d.novikov@imperial.ac.uk} 
\email{per.lilje@astro.uio.no} 

\altaffiltext{1}{Also at Centre of Mathematics for Applications,
University of Oslo, P.O.\ Box 1053 Blindern, N-0316 Oslo}
\altaffiltext{2}{Also at Jet Propulsion Laboratory, M/S 169/327, 4800 Oak Grove Drive,
  Pasadena CA 91109}
\altaffiltext{3}{Also at California Institute of Technology, Pasadena, CA
  91125}
\altaffiltext{4}{Also at Astrophysics Group, Imperial College London,
Blackett Laboratory, Prince Consort Road, London SW7~2AZ, UK (current address)}

\author{A.\ J.\ Banday}
\affil{Max-Planck-Institut f\"ur Astrophysik, Karl-Schwarzschild-Str.\
1, Postfach 1317, D-85741 Garching bei M\"unchen, Germany}
\email{banday@mpa-garching.mpg.de}

\and

\author{K.\ M.\ G\'orski} 
%\author{K.\ M.\ G\'orski\altaffilmark{3}} 
\affil{Jet Propulsion Laboratory, M/S
69/327, 4800 Oak Grove Drive, Pasadena CA 91109 \\ 
Warsaw University Observatory, Aleje
Ujazdowskie 4, 00-478 Warszawa, Poland}
%\altaffiltext{3}{Also at Warsaw University Observatory, Aleje
%Ujazdowskie 4, 00-478 Warszawa, Poland}
\email{krzysztof.m.gorski@jpl.nasa.gov}

%\date{Received 2004 January 14/ Accepted 2004 May 10}

\begin{abstract}
The three Minkowski functionals and the recently defined length of the
skeleton are estimated for the co-added first-year \emph{Wilkinson
Microwave Anisotropy Probe} (\emph{WMAP}) data
and compared with 5000 Monte Carlo simulations, based on Gaussian
fluctuations with the a-priori best-fit running-index power
spectrum and \emph{WMAP}-like beam and noise properties. Several
power spectrum-dependent quantities, such as the number of stationary
points, the total length of the skeleton, and a spectral parameter,
$\gamma$, are also estimated. While the area and length Minkowski
functionals and the length of the skeleton show no evidence for
departures from the Gaussian hypothesis, the northern hemisphere genus
has a $\chi^2$ that is large at the 95\% level for all scales. For the
particular smoothing scale of $3\fdg40$ FWHM it is larger than that
found in 99.5\% of the simulations. In addition, the \emph{WMAP} genus for
negative thresholds in the northern hemisphere has an amplitude that
is larger than in the simulations with a significance of more than
$3\,\sigma$.  On the smallest angular scales considered, the number of extrema
in the \emph{WMAP} data is high at the $3\,\sigma$ level. However, this
can probably be attributed to the effect of point sources.
Finally, the spectral parameter $\gamma$ is high at the 99\% level in the
northern Galactic hemisphere, while perfectly acceptable in the
southern hemisphere. The results provide strong evidence for the presence
of both non-Gaussian behavior and an unexpected power asymmetry between the
northern and southern hemispheres in the \emph{WMAP} data.
\end{abstract}

\keywords{cosmic microwave background -- cosmology: observations -- 
methods: statistical}

%\maketitle

\section{Introduction}

Over the last few years much interest has been focused on constraining
non-Gaussian contributions to the cosmic microwave background
(CMB). This interest is based on three facts: First, if the CMB
anisotropy field is in fact Gaussian, the power spectrum (or,
equivalently, the two-point correlation function) contains all the
statistical information required to characterize the field. Under the
Gaussian hypothesis, therefore, one may compress the statistical
information content of the full data set, consisting of several
millions of data points, into one single vector, $C_{\ell}$, of
perhaps a few thousand power spectrum coefficients. Second, most
current theories predict a Gaussian (or at least nearly Gaussian)
anisotropy field, and significant departures from Gaussianity could
point toward new physics (e.g., Wang \& Kamionkowski 2000; Lyth \&
Wands 2002; Bouchet et al. 2002; Bernardeau \& Uzan 2003). Finally,
residual foregrounds or systematic artifacts caused by the instrument or
data processing are likely to manifest themselves as non-Gaussian
signal contributions.  Tests for non-Gaussianity afford possible
methods for detecting, characterizing, and subsequently removing such
effects.  Further impetus for the study of such signatures has arisen
in recent months from putative detections of non-Gaussian features in
the \emph{Wilkinson Microwave An\-iso\-tro\-py Probe}
(\emph{WMAP}; Bennett et al.\ 2003a) first-year sky maps (Coles et al.\
2004; Naselsky et al.\ 2003; Vielva et al.\ 2004; Copi, Huterer, \&
Starkman 2004; Park 2004; Eriksen et al.\ 2004).

There is no unique way for a random field $\Delta T(\theta,\phi)$ to
manifest non-Gaussianity; hence, there is no generically superior,
high-sensitivity test for all of the possible ways in which a field
can be non-Gaussian.  In order to perform a thorough analysis, one
therefore has to apply a wide range of qualitatively different tests
that attack the problem from different directions.  Different classes
of tests, e.g., harmonic space methods such as the bispectrum and
trispectrum, the one-point distribution function (e.g., skewness and
kurtosis), $N$-point correlation functions in real space,
wavelet-based tests, and morphological measures such as those
described in this paper, all probe different aspects of deviation from
Gaussianity and are probably sensitive to different classes of such
behavior. While tests performed in harmonic space are most sensitive
to non-Gaussianities concentrated on characteristic angular scales as
quantified by $\ell$, tests performed in real space are most sensitive
to non-Gaussianities localized to specific regions on the sky.

In this paper we present estimates of four functions, all related to
the morphology of hot and cold areas, computed from the \emph{WMAP}
first-year data. These four functions are the three Minkowski
functionals \citep{minkowski} and the ``skeleton'' length (Novikov,
Colombi, \& Dor\'e 2003), which all are known for their ability to
measure non-Gaussian effects. Historically, cosmological tests based
on topology were probably introduced by \citet{Gott:1986} in the
context of the distribution of galaxies. The genus Minkowski
functional (or equivalently, the Euler-Poincar\'e characteristic) was
first suggested as a way of detecting the non-Gaussianity of the CMB
by \citet{coles_barrow} and \citet{coles:1988}. Topological tests of
the CMB anisotropies were developed further by, e.g., Gott et al.\
(1990), and finally, the full set of Minkowski functionals was
introduced into the field of CMB research by \citet{Schmalzing:1998},
\citet{Winitzki:1998}, and \citet{Novikov:1999}. These have recently
been joined by the skeleton length measure \citep{Novikov:2003}, which
naturally belongs to the same group of statistics.  One major
advantage of these tests is that they are relatively undemanding on
computer resources, and therefore it is feasible to apply these tests
to the megapixel CMB data sets from modern experiments, such as
\emph{WMAP}.

While the skeleton length has not been measured for the \emph{WMAP}
first-year data set to this date, at least three
other groups have computed the genus Minkowski functional for this
data set, namely, \citet{Komatsu:2003}, \citet{Colley:2003}, and
\citet{Park:2004}. While the former group does not describe their
algorithms in great depth, the latter two use pixel-by-pixel methods
for computing the genus. We choose a third method based on the first
and second derivatives of the map for finding all stationary points in
the map. Although each of the four functions has an explicit analytic
expectation value for a Gaussian field, we find it generally more
convenient to calibrate our results with Monte Carlo simulations in
order to assess the importance of the effects of realistic beam
profiles, non-uniform noise, and partial sky coverage.  As a by-product
of these functional measurements, we also determine a number of power
spectrum-dependent quantities, such as the number of extrema and the
total length of the skeleton. Each of these statistics may be regarded
as an independent test of the underlying power spectrum.

It should be noted that another recent analysis by
\citet{hansen:2004} computes the local curvature distributions of the
\emph{WMAP} data, using a technique very similar to ours. Although
they do not obtain quite as strong detections with this curvature
measure as determined here with the genus statistic, the two sets of
results are in excellent agreement.

The rest of the paper is organized as follows. In \S
\ref{sec:definitions} we review the definitions of the Minkowski
functionals and the skeleton, while \S \ref{sec:algorithms}
describes how to compute each of these functions from a pixelized
map. The necessary preparations and the statistical methods to be applied
to the actual \emph{WMAP} data are summarized in \S\S
\ref{sec:preparations} and \ref{sec:quantify}. Finally, the results
are shown in \S \ref{sec:results}, before comparing with other
results and making some final remarks in \S
\ref{sec:conclusions}.

\section{The Estimators}
\label{sec:definitions}

In our analysis, we study the normalized anisotropy field,
\begin{equation}
\nu(\phi, \theta) = \frac{\Delta T(\phi, \theta)}{\sigma_0},
\label{eq:norm_field}
\end{equation}
where the standard deviation $\sigma_0$ of the anisotropy field over
the observed region $\delta \Omega$ of area $A_{\textrm{obs}}$ is
defined by $\sigma_0^2 = \frac{1}{A_{\textrm{obs}}}\,\int_{\delta
\Omega} (\Delta T-\bigl<\Delta T\bigr>)^2 d\Omega$.

This normalization is necessary in order to minimize the impact of the
realization-dependent power spectrum -- we are more interested in
departures from Gaussianity, than in possible power spectrum
deviations.

For a study of the geometric properties of the CMB field to be
meaningful, we have to consider the \emph{shape} of connected
sets. This is done here, as usual, by studying \emph{excursion sets}
of the two-dimensional CMB-field on the celestial sphere, i.e., the
set consisting of the parts of the sky having a temperature above a
given threshold $\tilde{\nu}$. The main interest is in seeing how the
morphological descriptors vary as functions of $\tilde{\nu}$.

\subsection{The Minkowski functionals}

A well known theorem of integral geometry \citep{hadwiger} says that
under relatively unrestrictive conditions (that the descriptors are
rotationally and translationally invariant and that they preserve
additivity and convex continuity), the morphology of a convex body in
$N$-dimensional space can be completely described by $N+1$ Minkowski
functionals \citep{minkowski}. Therefore, a morphological description
of the CMB anisotropy field on the two-dimensional celestial sphere
can be obtained through three Minkowski functionals, and \emph{all}
possible morphological descriptors obeying the above conditions can be
expressed as a linear combination of those.  Thorough reviews of the
two-dimen\-sion\-al Minkowski functionals on a sphere are given by,
e.g., \citet{Schmalzing:1998} and \citet{Winitzki:1998}.

The first two-dimensional Minkowski functional we use is the
normalized area $\mathcal{A}$ of the excursion set $\mathcal{R}$,
i.e., the area of $\mathcal{R}$ as a fraction of the full area of the
survey. Using the HEALPix\footnote{Available from
http://www.eso.org/science/healpix} pixelization (G\'orski, Hivon, \&
Wandelt 1999), in which all pixels have exactly equal area, the area
functional is easily computed by counting the number of pixels with a
field value above the threshold under consideration,
\begin{equation}
  \mathcal{A}(\tilde{\nu}) = \frac{1}{A_{\textrm{obs}}} \int_{\mathcal{R}} dA \approx
  \frac{N_{\textrm{pix}}(\nu >
\tilde{\nu})}{N_{\textrm{pix}}^{\textrm{tot}}}.
\label{eq:areafunc}
\end{equation}

The second two-dimensional Minkowski functional is the length of the
border of the excursion set,
\begin{equation}
  \mathcal{L}(\tilde{\nu}) = \int_{\delta \mathcal{R}} d\ell.
\end{equation}
The details of computing this measure on a pixelized map are
provided in detail below. We follow the convention of normalizing the
function to have unit integral value,
\begin{equation}
  \int \mathcal{L}(\tilde{\nu}) d\tilde{\nu} = 1.
\end{equation}

The third Minkowski functional is the genus $\mathcal{G}$, which is
defined as the (normalized) number of connected hot spots (regions
above the threshold, i.e., connected parts of $\mathcal{R}$) minus the
number of connected cold spots (regions below the threshold, i.e.,
connected parts of the whole survey area minus
$\mathcal{R}$). However, here we approximate $\mathcal{G}$ with the
number of maxima plus the number of minima minus the number of saddle
points in the excursion set (Novikov, Schmalzing, \& Mukhanov 2000),
divided by the sum of \emph{all} stationary points in the whole
un-thresholded field,
\begin{equation}
  \mathcal{G}(\tilde{\nu}) = \frac{N_{\textrm{max}}(\tilde{\nu}) +
  N_{\textrm{min}}(\tilde{\nu}) - N_{\textrm{sad}}(\tilde{\nu})}
{N_{\textrm{max}}(-\infty) +
  N_{\textrm{min}}(-\infty) + N_{\textrm{sad}}(-\infty)}.
\end{equation}

\subsection{The skeleton length}

The skeleton length was recently introduced as a diagnostic for
Gaussianity by \citet{Novikov:2003}. For a random field on ${\cal{\bf
R}}^2$, it is defined as the zero-contour line of the map
\begin{equation}
\mathcal{S} = T_x T_y (T_{xx} - T_{yy}) + T_{xy} (T_y^2 - T_x^2),
\label{eq:S_map_def}
\end{equation}
where $T_x$ and $T_y$ denote the first derivatives of the random
field in two orthogonal directions and the $T_{ij}$-values denote the second
derivatives. 

\begin{figure*}
\plotone{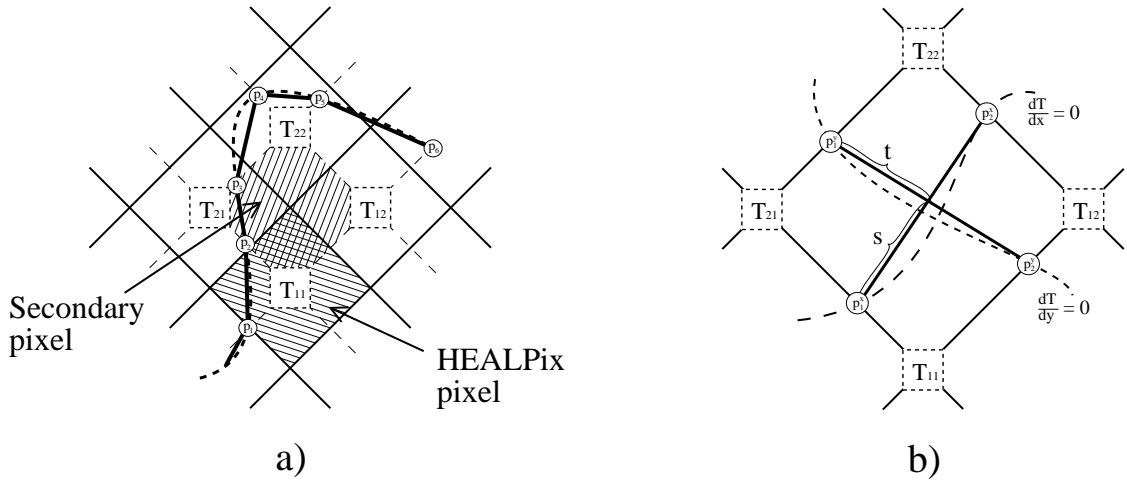}
\caption{Tracing contour lines and localizing stationary points by
means of linear interpolation. (\emph{a}) Temperature values $T_{ij}$
are known only at the centers of the HEALPix pixels, and linear
interpolation is therefore used to approximate the true contour
line. The true contour line is shown as a dashed curve in this figure,
and the linear approximation as solid line segments. In this process
it is useful to construct a set of secondary pixels defined by letting
the centers of the HEALPix pixels define their vertices. In order to
trace the contour line, one then has to check all such secondary
pixels, searching for edges for which the two vertices have different
signs relative to the contour line. Once such an edge has been
located, the point of intersection (marked as $p_i$ in the figure) is
approximated by linear interpolation, according to Equation
(\ref{eq:crossing}). (\emph{b}) Localizing stationary points is
similar to tracing the contour lines. In this case one focuses on the
zero-contours of the first derivative maps and searches for
secondary pixels in which both derivatives are zero. Once such a pixel
is found, the position of intersection is approximated by linear
interpolation, i.e., by solving Equation (\ref{eq:localizing_extrema})
for $s$ and $t$. If $0<s,t<1$, then the point of intersection lies
within the secondary pixel under consideration.}
\label{fig:contour_trace}
\end{figure*}

When applying the above expression to a field on the sphere, it is
crucial to notice that $\mathcal{S}$ is a rotationally invariant
quantity. Therefore, we may construct a local coordinate system at each
point and compute the derivatives in that local system before adding
them together. In particular, we may choose our coordinate system to
be the standard latitude-longitude system on the sphere, and the
derivatives may be chosen to be the covariant derivatives. The
skeleton map $\mathcal{S}$ on the sphere may therefore be expressed by
\begin{equation}
\mathcal{S} = T_{;\phi} T_{;\theta} (T_{;\phi\phi} -
T_{;\theta\theta}) + T_{;\phi\theta} (T_{;\theta}^2 -
T_{;\phi}^2).
\label{eq:S_sphere}
\end{equation}
where the semicolons as usual denote covariant derivatives. We
later discuss how to compute them.

Although Equation (\ref{eq:S_sphere}) may be difficult to interpret,
its geometric interpretation is intuitive: the zero-contour lines of
$\mathcal{S}$ are the set of lines that extend from extremum to
extremum along lines of maximum or minimum gradient, and the set of
all such lines is collectively called the skeleton of the field.

The statistic we are interested in is the length of the skeleton of
that part of the field that lies above some threshold $\tilde{\nu}$,
normalized by the length of the skeleton of the whole un-thresholded
field. Obviously this can be computed by the same algorithm
as is used for computing the length Minkowski functional
$\mathcal{L}$, since the skeleton is the zero-contour of the
$\mathcal{S}$-map.

For a Gaussian field \citep{Novikov:2003}, the differential skeleton
length depends only on one spectral parameter, $\gamma$, defined as
\begin{equation}
\gamma \equiv \frac{\sigma_1^2}{\sigma_0 \sigma_2},
\end{equation}
where $\sigma_0^2$ is the variance of the map, $\sigma_1^2$ is the
variance of the first-order derivatives, and $\sigma_2^2$ is the variance
of the second-order derivatives. We later use this quantity as an
independent statistic when studying the \emph{WMAP} data.

Actually, it has been shown that for a Gaussian field, the
differential skeleton length is
\begin{equation}
L(\tilde\nu,\gamma)=\frac{1}{\sqrt{2\pi}}e^{-\tilde\nu^2/2}[1+0.17\gamma^2(1-
\tilde\nu^2)].
\end{equation}

\section{Algorithms}
\label{sec:algorithms}

As seen from the definitions, in addition to just counting pixels for
the area Minkowski functional (Equation \ref{eq:areafunc}), we need
two distinct algorithms for computing the four functions (i.e., the
three Minkowski functionals and the length of the skeleton). For the
skeleton length and for the length Minkowski functional it is necessary 
to estimate the length of a contour line, and for the genus functional
the positions of all stationary points must be located and 
the field values at these points estimated.

Both these operations, as well as the determination of the skeleton,
depend on derivatives of the field. Such derivatives are
well-behaved only if the map itself is smooth (that is, not dominated
by pixel noise), and we therefore consider it prudent to filter each
map before applying our algorithms. In addition, simple experiments
show that robust estimation of the contour line lengths requires a
high pixel resolution compared to the smoothing scale: adequate
smoothing is then an essential step in the analysis. A brief
discussion of this and related topics is given in
\S\ref{sec:preparations}. As pointed out by, e.g.,
\citet{Winitzki:1998}, numerical techniques for estimating the length
of a contour line or positions of stationary points on real CMB maps
with noise, pixelization, masks, etc., have to be chosen with
care. Our experiments show that the algorithms employed here are
sufficiently insensitive to such effects and give accurate results.

\subsection{Measuring the length of an iso-contour line}

Our starting point is thus a smoothed, pixelized map in which we know
the field values at the pixel centers. We want to trace the underlying
contour lines between those pixel centers and estimate their
lengths. The method for doing so is based on linear interpolation and
is almost identical to the methods adopted by, e.g.,
\citet{shandarin:2002} and \citet{Novikov:1999}. However, we also use
this method for locating the stationary points; therefore, we
review the algorithm in detail here.

We start by constructing a secondary set of pixels, defined by letting
the original pixel centers be at the corners of the new, secondary
pixels. The secondary pixels are then checked to determine those that
are crossed by a contour line. These are obviously the secondary
pixels where at least one vertex has a lower value and at least one
vertex has a higher value than the contour line. This process is
illustrated in the left-hand panel of Figure \ref{fig:contour_trace}.

Once such a ``crossed'' secondary pixel has been found, it is easy
to determine through which edges the contour line passes; a contour line
crosses an edge if one vertex has a field value larger than the
contour value and the other vertex has a smaller value. The point on
the edge where the contour line crosses is estimated by a linear
approximation,
\begin{equation}
\mathbf{n}_{\textrm{cont}} = \biggl|\frac{\Delta T_2 - \Delta
  T_{\textrm{cont}}}{\Delta T_2 - \Delta T_1}\biggr| \mathbf{n}_1 +
  \biggl|\frac{\Delta T_1 - \Delta T_{\textrm{cont}}}{\Delta T_2 -
  \Delta T_1}\biggr| \mathbf{n}_2.
\label{eq:crossing}
\end{equation}
Here $\mathbf{n}_{\textrm{cont}}$ (given as an unnormalized vector
from the center of the sphere) is the point at which the contour line crosses the
secondary pixel edge, $\Delta T_1$ and $\Delta T_2$ are the
field values at the two vertices with corresponding positions $\mathbf{n}_1$
and $\mathbf{n}_2$, while $\Delta T_{\textrm{cont}}$ is the field value
on the contour line.

After locating both points on the edges of the secondary pixel where
the contour line enters and exits the pixel, the angular length of
the line segment within that pixel is estimated by computing the dot
product of the two vectors,
\begin{equation}
  \delta \ell = \arccos(\hat{\mathbf{n}}_{\textrm{cont}}^1 \cdot
  \hat{\mathbf{n}}_{\textrm{cont}}^2),
\end{equation}
where $\hat{\mathbf{n}}_{\textrm{cont}}^i$ are the {\it normalized}
unit vectors found in Equation (\ref{eq:crossing}).

\begin{figure*}
\plotone{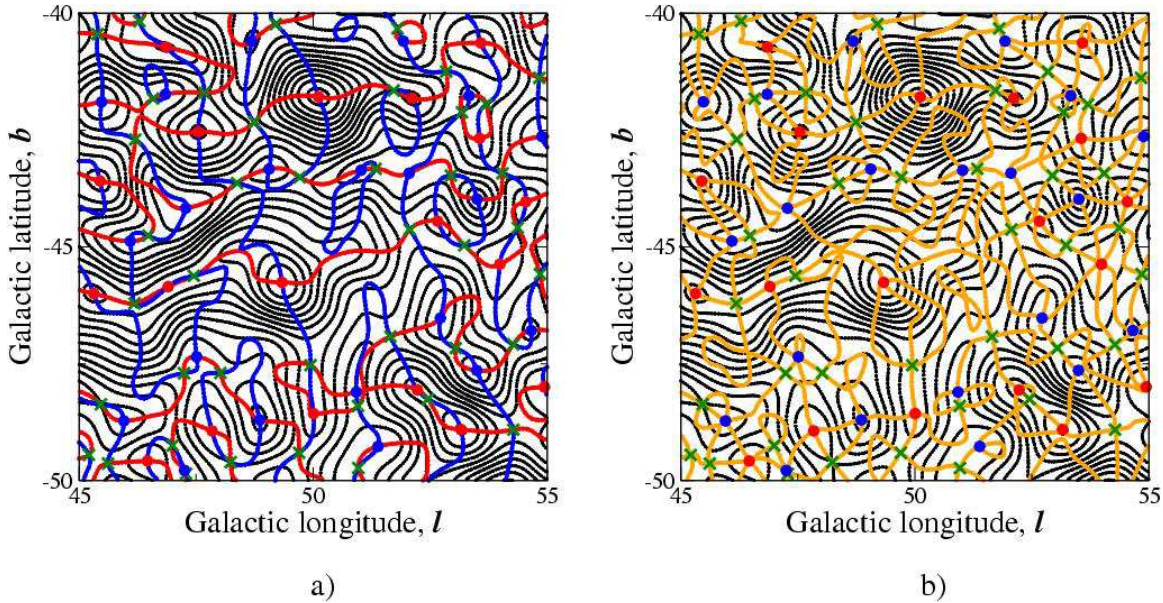}
\caption{A $10^{\circ} \times 10^{\circ}$ patch from the CMB sky as
  measured by \emph{WMAP}, with (\emph{a}) $dT/d\theta = 0$ marked in blue and
  $dT/d\phi = 0$ marked in red, and (\emph{b}) the skeleton of the field
  marked in yellow. In both figures red filled circles indicate maxima, blue
  filled circles indicate minima and green crosses indicate saddle points.}
\label{fig:patch}
\end{figure*}

In those cases in which more than two edges of a pixel are crossed by a
contour line, we assume that two contour lines cross each other inside
the pixel. The justification for this assumption is simply that we are
interested only in the total length of the contour line, not its
shape, and the possible error introduced in the total length by this
assumption is obviously very small because of the small pixel size.

\subsection{Locating the stationary points of a random field on a sphere}

In order to compute the genus Minkowski functional, we need to locate
all stationary points -- both minima, maxima, and saddle points. While
extrema alone may be found by searching for pixels that have neighbors
with all higher or lower temperature values (as is implemented in the
HEALPix hotspot facility), this method cannot be used for locating
saddle points. Therefore, a different approach is adopted in this
work, namely, a method based on the covariant derivatives of the
random field.

A stationary point has by definition vanishing first partial
derivatives computed in two orthogonal directions. The particular type
of stationary point is determined by the eigenvalues of the Hessian
matrix: if all eigenvalues are positive, the point is a minimum; if
all are negative, the point is a maximum; and if there are both
positive and negative eigenvalues, it is a saddle point.  To find the
positions and types of all stationary points, we therefore need to be
able to efficiently compute the first and second derivatives of the
temperature field.  We use the covariant derivatives in the standard
polar coordinate system,
\begin{align}
\Delta T_{;\phi} &= \frac{1}{\sin \theta} \frac{\partial \Delta T}{\partial \phi}, \\
\Delta T_{;\theta} &= \frac{\partial \Delta T}{\partial \theta}.
\label{eq:covar_deriv}
\end{align}
In this paper, these derivatives are only used to determine the skeleton
and to locate the stationary points, but, as shown by
\citet{Schmalzing:1998}, it is possible to express all the Minkowski
functionals in terms of the field itself and its first- and
second-order derivatives.

It is most convenient not to compute the covariant derivatives of the
temperature anisotropy field directly from the pixel values, but
rather to do it from the expansion of the anisotropy field in
spherical harmonics,
\begin{align}
\Delta T(\theta, \phi) &= \sum_{l=0}^{l_{\textrm{max}}}
\sum_{m=-l}^{l} a_{lm} \: Y_{lm}(\theta, \phi) \\
&=  \sum_{l=0}^{l_{\textrm{max}}}
\sum_{m=-l}^{l} a_{lm} \: X_{lm} P_{lm}(\cos \theta) e^{im\phi},
\end{align}
where the partial derivatives are found term by term. Here $X_{lm}
\equiv \left\{\left[(2l+1)/4\pi\right]
\left[(l-m)!/(l+m)!\right]\right\}^{1/2}$ is a
normalization constant.

By using the identity for associated Legendre polynomials,
\begin{equation}
\frac{d P^m_l}{d x} = \frac{1}{1-x^2} \biggl[
  \frac{(l+1)(l+m)}{2l+1} P^m_{l-1} - \frac{l(l-m+1)}{2l+1}
  P^m_{l+1}\biggr],
\label{eq:deriv_plm}
\end{equation}
where $x = \cos \theta$, one
finds that the covariant derivatives are given by
\scriptsize
\begin{align*}
\Delta T_{;\phi} &= \frac{i}{\sin \theta}\sum_{l=0}^{l_{\textrm{max}}} \sum_{m=-l}^{l}
  (m a_{lm}) Y_{lm} \\
\Delta T_{;\theta} &= \frac{1}{\sin\theta} \biggl[\sum_{l=2}^{l_{\textrm{max}}+1} \sum_{m=-l+1}^{l-1}
\frac{(l-1)(l-m)}{2l-1} a_{l-1\,m} \frac{X_{l-1\,m}}{X_{lm}} Y_{lm} - \\ 
& \quad \quad\quad\quad-\sum_{l=0}^{l_{\textrm{max}}-1} \sum_{m=-l}^{l}
 \frac{(l+2)(l+m+1)}{2l+3} a_{l+1\,m} \frac{X_{l+1\,m}}{X_{lm}} Y_{lm}\biggr].
\end{align*}
\normalsize
These derivatives may then be computed very efficiently using the
HEALPix routines map2alm and alm2map.

In principle, one could go through this process once more to find
equivalent expressions for the second derivatives. However, in
practice one obtains numerically more stable results by making new
pixelized maps of the first derivatives of the original map, expanding
those in spherical harmonics using HEALPix routines, and then applying
the above formulae to them, rather than using second-order
expressions on the temperature map itself. The cost for this extra
stability comes in the form of two extra spherical harmonics
transforms, but the extra CPU time required is acceptable.  By
repeated application of the above formulae we thus find all first and
second derivatives of the anisotropy field in all pixel centers.

The next problem is to find the points of intersection of the lines
$\Delta T_{;\phi} = 0$ and $\Delta T_{;\theta} = 0$, since the
stationary points must be at these intersections. This is done with an
algorithm similar to that described for tracing the iso-contour lines;
once again we focus on the set of secondary pixels, but this time we
search for secondary pixels being crossed by both  ``iso-contour
lines'' $\Delta T_{;\phi} = 0$ and $\Delta T_{;\theta} = 0$.

Once such a secondary pixel is found where the two
zero-derivative curves cross its edges, it must be determined whether the two
lines actually cross each other \emph{inside} that pixel. We find the
position of all four edge crossings (two for each contour line) as for
Equation (\ref{eq:crossing}), project these positions into a local,
two-dimensional coordinate system centered on the pixel, and solve the
system
\begin{equation}
s \,\mathbf{p}_{\textrm{in}}^{\phi} + (1-s)
\mathbf{p}_{\textrm{out}}^{\phi} = t
\,\mathbf{p}_{\textrm{in}}^{\theta} + (1-t)
\mathbf{p}_{\textrm{out}}^{\theta},
\label{eq:localizing_extrema}
\end{equation}
of linear equations.  Here the four $\mathbf{p}_{\textrm{in/out}}^{i}$
are four two-dimensional vectors representing each of the four 
crossing-points of the edge of the pixel.

If $0 \le s, t \le 1$, the two lines do intersect each other inside
the pixel, and the point of intersection is estimated by substituting
either $s$ or $t$ into the respective side of the above
equation. Figure \ref{fig:contour_trace}(b) illustrates this method.

After locating such a stationary point, one must then estimate both
the field value (for thresholding) and the second derivatives at that
point. For this purpose we adopt a weighting method that takes into
account the values at all four vertices. First, the angular
distance between the point of intersection and the four vertices
is computed,
\begin{equation}
u_{i} = \arccos (\mathbf{p} \cdot \mathbf{p}_i).
\end{equation}
The relative weights, $\tilde{w}_i$, are then evaluated as the smallest
angular distance divided by $u_i$,
\begin{equation}
\tilde{w}_i = \frac{\min_j u_j}{u_i}.
\end{equation}
Finally, the true weights, $w_i$, are found by normalizing the sum to
unity, $\sum w_i = 1$.

The field value and the second derivatives at the point of
intersection may now be estimated as weighted sums over the vertices,
\begin{align}
\Delta T &= \sum_{i = 1}^{4} w_i \Delta T^i\\
\Delta T_{;jk} &= \sum_{i = 1}^{4} w_i \Delta T^i_{;jk}.
\end{align}

The last step in this process is to determine the type of stationary
point. This is done from the characteristic equation,
\begin{equation}
\lambda^2 - (\Delta T_{;\theta\theta} + \Delta T_{;\phi\phi}) \,
\lambda + ( \Delta T_{;\theta\theta}\Delta T_{;\phi\phi} - \Delta
T_{;\theta\phi}^2 ) = 0.
\end{equation}
The stationary point is a maximum if the solutions of the
characteristic equation $\lambda_1, \lambda_2 < 0$, a minimum if
$\lambda_1, \lambda_2 > 0$ and a saddle point if $\lambda_1$ and
$\lambda_2$ have different signs (assuming that the point is
non-degenerate, i.e., that $\lambda_i \ne 0$ which is always true for
real-world fields).

The procedure is illustrated in Figure (\ref{fig:patch}). A contour
map of a patch of the CMB sky is shown in the left panel.  The $\Delta
T_{;\theta}=0$ contours are the blue lines and the $\Delta
T_{;\phi}=0$ contours are the red lines. The stationary points
(maxima: \emph{red filled circles}; minima: \emph{blue filled
circles}; saddle points: \emph{green crosses})
are seen to be at the intersections of the blue and red lines. In the
right-hand panel the skeleton of the field is overlaid on the same map as
a yellow contour line.

\section{The \emph{WMAP} data and the simulations}
\label{sec:preparations}

The aim of this paper is to measure the three Minkowski functionals
and the length of the skeleton as a function of the threshold height
for the
\emph{WMAP} first-year maps and to compare these 
to the same quantities estimated from a Monte Carlo set of 5000
simulated Gaussian sky maps.  The Monte Carlo ensemble is based on
the best-fit
\emph{WMAP} power spectrum with a running index, which may be
down-loaded from the LAMBDA\footnote{http://lambda.gsfc.nasa.gov} Web
site.

\begin{figure}
\mbox{\epsfig{file=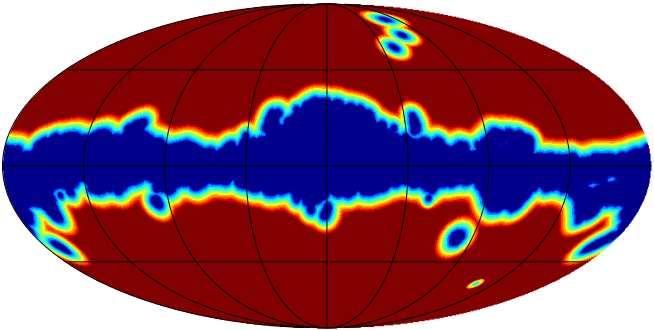,height=88mm,angle=90}}
\caption{Masks used in the computations of the Minkowski
  functionals and the skeleton for FWHMs between $0\fdg53$ and
  $4\fdg26$. The masked regions at high latitudes are
extended ``semi-point'' sources excluded by the Kp0 mask.} 
\label{fig:masks}
\end{figure}

Our study is based on the \emph{co-added} \emph{WMAP} map (Hinshaw et
al.\ 2003), and in order to produce the best possible approximations
to this map, the simulations are processed in exactly the same way as
the \emph{WMAP} map. In short, each realization is constructed as a
weighted sum over each of the cosmologically important \emph{WMAP}
channels (Q1, Q2, V1, V2, W1, W2, W3, and W4), taking into account the
particular beam and noise properties of each channel
\citep{Spergel:2003}. We use foreground-corrected maps produced by the
template fitting method of \citet{Bennett:2003b} to suppress foreground 
contamination. 

The publicly-released maps are available at the LAMBDA Web-site.
However, a problem was recently discovered with these maps (see the
LAMBDA web
page\footnote{http://lambda.gsfc.nasa.gov/product/map/IMaps\_cleaned.cfm},
where a more thorough description is given); the templates used in
this process were convolved with a too low $l_{\textrm{max}}$ (the
maximum multipole component), resulting in ringing around strong point
sources. For this reason we have also generated our own templates and
performed the analysis for these maps as well. The differences between
the two sets of results are insignificant on the scales we consider;
therefore, we choose to report only the results for the official
maps in what follows.

For both the \emph{WMAP} data and the simulated maps, a number of data
processing steps are performed to minimize the impact of residual
foregrounds, the applied Galaxy cut, pixel noise, etc.  First, the
\emph{WMAP} Kp0 mask with point sources included (i.e., we do not
exclude the 700 point sources present in the mask specified by the
\emph{WMAP} team) is defined as the base mask.  The decision not to
exclude the 700 point sources is based on two factors. First, we are
mainly interested in scales larger than $2^{\circ}$, and for such
large scales the impact of point sources is small (see, e.g., Spergel
et al.\ 2003). Later in the paper the impact of point sources is
explicitly tested by applying a median filter to the map. Second,
since the maps are smoothed, we also have to expand the mask
accordingly. If the 700 point sources actually were excluded by the
base mask, the \emph{expanded} mask would accept only a very small
fraction of the sky. It is therefore better to include the point
sources in the analysis, and then later on determine from the results
themselves whether this is justified or not.  To test for any effects
of the exact shape of the mask, we also perform our analysis using a
base mask with the additional constraint $\vert b\vert < 30\degr$.

The next processing step is to remove the best-fit mono- and dipoles
from the maps, with coefficients computed from the accepted region
only. Then, the masked region is nullified, and the mask boundary is
apodized, before the maps are smoothed with a Gaussian beam of given
FWHM (which will be varied, see below). This smoothing operation
consists of a spherical harmonics transform of the original map,
followed by a multiplication with the Legendre transform of the
Gaussian beam. Then the smoothed map is constructed through an inverse
spherical harmonics transform. At this point we keep the final
multipole expansion components, $a_{lm}$, for computing the
derivatives (see \S\ref{sec:algorithms}) at a later stage.

The next step is to expand the excluded regions of the mask in all
directions as a safeguard against border effects. This is done
according to the procedures of \citet{Schmalzing:1998}; the mask map
consisting of zeros and ones is convolved with a Gaussian beam of the
required FWHM, and the new, expanded mask is then determined by
including all pixels with values higher than some given threshold, for
instance, 0.95 or 0.99. For maximum safety we perform this operation
\emph{twice}, each time with a limiting threshold of 0.99. The
resulting masks are shown in Figure \ref{fig:masks}.  Finally, the map
is normalized according to Equation \ref{eq:norm_field}, including
pixels in the accepted region only.

The analysis is carried out for a large number of smoothing FWHMs in
order to probe different angular scales. The beam widths selected were
$0\fdg53$, $0\fdg64$, $0\fdg85$, $1\fdg28$, $1\fdg70$, $2\fdg13$,
$2\fdg55$, $2\fdg98$, $3\fdg40$, $3\fdg83$, and $4\fdg26$
FWHM\footnote{After having completed the computations, we discovered
an error in a computer code that effectively multiplied all smoothing
FWHMs by $\sqrt{8\,\log 2} = 2.35$. Thus, the FWHMs that originally
had been chosen to be multiples of $1\degr$ were in reality multiples
of $0\fdg43$, causing the somewhat unnatural looking smoothing scales
in this paper.}. Note that this smoothing is applied directly to the
observed maps; therefore, the Gaussian smoothing beam is in
addition to the experimental \emph{WMAP} beams. In total, the narrowest of
the beams ($0\fdg53$ FWHM) is sensitive to multipoles up to about
$\ell \approx 600$; therefore, both noise and point sources are
expected to play a major role at this and smaller scales. Here it is
worth noticing that \citet{Park:2004} considered even smaller scales,
since the maps were not explicitly smoothed at all, except for a small
effect introduced by a stereo-graphical projection. However, in the
subsequent analysis, the 700 point sources resolved by \emph{WMAP}
were masked out, and it was therefore possible to produce results on
smaller scales than presented in this paper.

The maximum multipole component, $\ell_{\textrm{max}}$, in each run is
chosen to match the corresponding FWHM. In particular, the values of
$\ell_{\textrm{max}}$ we choose are 650, 600, 550, and 450 for FWHMs
between $0\fdg53$ and $1\fdg28$ and 350 for the larger beams. The
HEALPix resolution parameter used internally in the computations is
$N_{\textrm{side}} = 1024$. By increasing $N_{\textrm{side}}$ from 512
to 1024, the total skeleton length is increased by a few percent for
the narrowest beam. However, by increasing $N_{\textrm{side}}$ from
1024 to 2048, the total skeleton length is increased only by a
negligible amount.  Thus, $N_{\textrm{side}} = 1024$ is the lowest
resolution able to support \emph{all} our analyses, although lower
resolutions could have been chosen individually for the wider beams.

The four statistics are estimated for 200 values of the threshold
height between $-4\,\sigma_0$ and $4\,\sigma_0$, but only a subset of
these values are used in each of the subsequent analyses.  Each
statistic is evaluated independently on the northern and southern
Galactic hemispheres. Interesting hemisphere effects have been
reported by several authors (Eriksen et al.\ 2004; Park 2004), and we
focus our tests accordingly.

\section{Quantifying the degree of agreement between simulations
  and observations}
\label{sec:quantify}

In order to quantify the degree of agreement between the simulations
and the observations, we adopt a diagonal $\chi^2$ statistic of the
form
\begin{equation}
\chi^2 = \sum_{\tilde\nu \in (-3,3)} \biggl[\frac{f(\tilde\nu) -
\bigl<f(\tilde\nu)\bigr>}{\sigma(\tilde\nu)}\biggr]^2,
\end{equation}
where $\tilde\nu$ is the threshold level, $\bigl<f(\tilde\nu)\bigr>$
is the average of the simulations, and $\sigma(\tilde\nu)$ is the
standard deviation of the measure under consideration. This statistic
is computed for both the simulations and the observed functions, and
the fraction of simulations with $\chi^2$ value \emph{lower} than the
$\chi^2$ determined from the real \emph{WMAP} map is connected to the
confidence level at which to either accept or reject the
null-hypothesis -- that the observed field is a natural member of the
Gaussian Monte Carlo ensemble.

Note that we choose a diagonal $\chi^2$ statistic for the functions in
this paper. In our experience, the ordinary $\chi^2$ statistic, which
takes into account bin-to-bin correlations through the covariance
matrix, behaves rather badly if the functions under study are binned
with a narrow bin size, as is the case here. If neighboring bins are
strongly correlated, the covariance matrix converges very slowly, and
in the study here, 5000 simulations would not be sufficient to produce
robust results. Furthermore, when the bin-to-bin correlations are
strong, the off-diagonal terms become more and more dominant, and the
shape of the inverse covariance matrix resembles that of a Mexican
hat. A Mexican hat matrix is the same as a high-pass
filter. Therefore, if the bin-to-bin correlations are strong, the
covariance matrix $\chi^2$ statistic is more sensitive to
\emph{fluctuations} in the functionals, rather than absolute
deviations. In some cases, one actually finds that the covariance
matrix $\chi^2$ statistic accepts functions that should obviously be
rejected by eye.

However, if the proposed model for the simulations is an adequate
description of the real data, then \emph{no} statistic should be able
to distinguish between the simulations and the observations. The
choice of statistic is therefore only a question of what one wants to
measure.

For the genus we adopt the machinery of \citet{Park:2004} and
perform a parametric fit of the form
\begin{equation}
\mathcal{G}_{\textrm{fit}}(\tilde\nu) = A (\tilde\nu-\Delta \tilde\nu)
e^{-\frac{1}{2}(\tilde\nu-\Delta\tilde\nu)^2},
\label{eq:genus_par}
\end{equation}
where $A$ and $\Delta\tilde\nu$ are free parameters. For a Gaussian
field it is expected that $\Delta\tilde\nu=0$, while $A$ is a normalization
constant depending on the power spectrum. This fit is performed over
three different ranges in $\tilde\nu$, namely, $\tilde\nu = (-3,3)$,
$(-2.5,-0.2)$, and $(0.2,2.5)$. The corresponding amplitudes are named
$A$, $A_{-}$, and $A_{+}$ and indicate, respectively, the best-fit
total amplitude and negative and positive amplitudes. We also define
an asymmetry parameter by
\begin{equation}
\Delta g = \frac{A_{-} - A_{+}}{A_{-} + A_{+}}.
\label{eq:genus_asym}
\end{equation}
For a Gaussian field there should be no asymmetry in the shape of the
genus, and consequently $\Delta g$ should also be zero. Note that this
definition of $\Delta g$ differs slightly from that of \citet{Park:2004}.

\begin{figure*}
\plotone{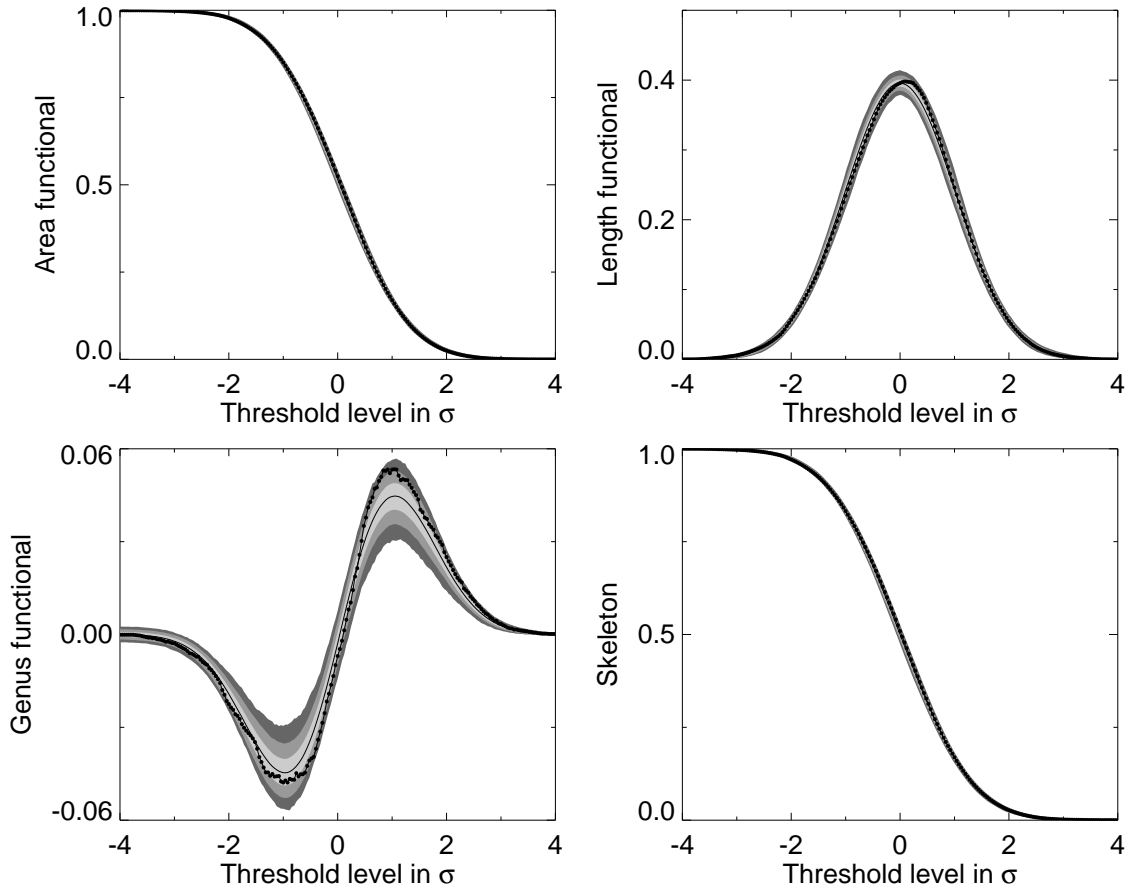}
\caption{Minkowski functionals and differential skeleton measured
  from the Kp0 sky, smoothed with a $1\fdg28$ FWHM Gaussian
  beam. Gray bands indicate 1, 2, and $3\,\sigma$ bands, and the solid line
  indicates the ensemble average. Filled circles show the observed functions.}
\label{fig:mink_full}
\end{figure*}

The fit of $\mathcal{G}_{\textrm{fit}}(\tilde\nu)$ to the data is
implemented as a non-linear search (using NAG\footnote{Numerical
Algorithms Group, http://www.nag.co.uk} routines), which
minimizes the functional
\begin{equation}
\mathcal{F} = \sum_{\tilde{\nu}} w(\tilde{\nu}) \:
\biggl[\mathcal{G}_{\textrm{obs}}(\tilde{\nu}) -
  \mathcal{G}_{\textrm{fit}}(\tilde{\nu})\biggr]^2,
\end{equation}
where $w(\tilde{\nu})$ is a weight factor. In this paper we choose
inverse variance weighting, $w(\tilde{\nu}) =
[\sigma^2_{\mathcal{G}}(\tilde{\nu})]^{-1}$, where
$\sigma^2_{\mathcal{G}}(\tilde{\nu})$ is the simulated variance of the
particular genus bin. However, almost all results are practically
independent of this particular choice, with the notable exception of
the total range amplitude $A$ for realizations with a large asymmetry
parameter $\Delta g$. In those cases, one may find that the amplitude
depends, but only weakly, on the weighting scheme. Moreover, we are mostly
concerned with the negative and positive threshold amplitudes and the
asymmetry parameter itself, and these quantities are very robust
against different weighting schemes.

\section{Results}
\label{sec:results}

\begin{deluxetable*}{crrrrr}
\tablewidth{0pt} 
\tabletypesize{\small} 
\tablecaption{Power spectrum
dependent measurements, Kp0 mask\label{tab:measurements}}
\tablecolumns{6}
\tablehead{FWHM & $\gamma$\quad\quad\quad$\:$ & Length\quad\quad\quad &
$N_{\textrm{min}}$ \quad\quad$\:$& 
$N_{\textrm{max}}$ \quad\quad$\:$& 
$N_{\textrm{sad}}$ \quad\quad$\:$}
\startdata
\cutinhead{Full sky}
 $0\fdg53$	        & 0.497 (95.2\%) & \phn\phm{}1870. (84.3\%) & 11\,258
(98.4\%) & 11\,330 (99.9\%) & 22\,565 (99.9\%) \\
 $0\fdg64$	        & 0.505 (93.9\%) & \phn\phm{}1584. (93.4\%) & 8558
(99.5\%) & 8515 (96.2\%) & 17\,049 (99.3\%) \\
 $0\fdg85$	& 0.503 (93.6\%) & \phn\phm{}1267. (93.0\%) & 5723
(91.8\%) & 5701 (81.7\%) & 11\,422 (92.2\%) \\
 $1\fdg28$	& 0.478 (93.4\%) & 904.6 (67.9\%) & 2939
(25.9\%) & 3019 (97.5\%) & 5958 (80.5\%) \\
 $1\fdg70$	& 0.468 (93.4\%) & 677.8 (58.6\%) & 1648
(\phn6.5\%) & 1716 (88.9\%) & 3352 (32.9\%) \\
 $2\fdg13$	& 0.465 (93.9\%) & 530.1 (84.0\%) & 1042
(49.8\%) & 1063 (85.9\%) & 2100 (71.3\%) \\
 $2\fdg55$	& 0.470 (95.0\%) & 429.7 (97.6\%) & 691
(48.2\%) & 705 (80.7\%) & 1395 (70.0\%) \\
 $2\fdg98$      & 0.478 (96.4\%) & 354.2 (85.7\%) & 494
(73.4\%) & 499 (84.7\%) & 986 (78.7\%) \\
 $3\fdg40$	& 0.488 (97.2\%) & 297.0 (56.0\%) & 366
(82.5\%) & 348 (28.5\%) & 719 (72.0\%) \\
 $3\fdg83$	& 0.496 (97.7\%) & 254.4 (52.0\%) & 285
(94.9\%) & 264 (35.3\%) & 554 (87.7\%) \\
 $4\fdg26$	        & 0.505 (98.0\%) & 221.2 (54.1\%) & 216
(80.5\%) & 204 (34.0\%) & 422 (70.1\%) \\

\cutinhead{Northern hemisphere}
 $0\fdg85$	& 0.535 (98.7\%) & 634.5 (70.3\%) & 2851
(62.4\%) & 2854 (65.9\%) & 5694 (60.4\%) \\
 $1\fdg28$	& 0.517 (98.9\%) & 454.8 (60.6\%) & 1493
(57.6\%) & 1537 (98.6\%) & 3023 (91.2\%) \\
 $1\fdg70$      & 0.511 (98.9\%) & 334.0 (23.9\%) & 811
(\phn1.3\%) & 875 (92.7\%) & 1681 (27.2\%) \\
 $2\fdg13$	& 0.514 (99.0\%) & 268.2 (83.6\%) & 524
(41.9\%) & 545 (90.5\%) & 1068 (75.9\%) \\
$2\fdg55$             & 0.526 (99.3\%) & 217.2 (89.3\%) & 339
(16.3\%) & 360 (80.9\%) & 702 (56.9\%) \\
 $2\fdg98$	& 0.537 (99.5\%) & 177.3 (35.5\%) & 245
(45.0\%) & 243 (37.0\%) & 490 (47.6\%) \\
 $3\fdg40$	& 0.548 (99.6\%) & 149.1 (22.6\%) & 181
(52.7\%) & 170 (10.2\%) & 356 (40.0\%) \\
 $3\fdg83$	& 0.555 (99.3\%) & 128.6 (33.3\%) & 143
(79.5\%) & 138 (57.5\%) & 284 (83.7\%) \\
 $4\fdg26$	        & 0.559 (98.9\%) & 113.0 (61.7\%) & 112
(80.6\%) & 105 (42.6\%) & 218 (72.6\%) \\

\cutinhead{Southern hemisphere}
 $0\fdg85$      & 0.492 (27.9\%) & 632.5 (94.3\%) & 2871
(94.2\%) & 2851 (83.0\%) & 5719 (94.0\%) \\
 $1\fdg28$      & 0.463 (25.5\%) & 449.8 (64.4\%) & 1453
(19.8\%) & 1477 (59.9\%) & 2937 (44.7\%) \\
 $1\fdg70$      & 0.449 (25.0\%) & 337.7 (83.2\%) & 834
(45.0\%) & 832 (38.8\%) & 1661 (34.2\%) \\
 $2\fdg13$      & 0.444 (25.0\%) & 261.9 (65.5\%) & 517
(52.4\%) & 517 (51.6\%) & 1029 (45.8\%) \\
 $2\fdg55$      & 0.445 (26.9\%) & 212.6 (91.9\%) & 346
(64.1\%) & 345 (58.6\%) & 688 (59.6\%) \\
 $2\fdg98$      & 0.452 (31.2\%) & 177.2 (97.4\%) & 252
(89.3\%) & 259 (97.9\%) & 501 (92.3\%) \\
 $3\fdg40$      & 0.460 (36.4\%) & 147.6 (77.5\%) & 187
(91.6\%) & 175 (48.3\%) & 365 (86.5\%) \\
 $3\fdg83$      & 0.471 (42.9\%) & 125.7 (63.1\%) & 139
(82.3\%) & 127 (23.7\%) & 268 (65.5\%) \\
 $4\fdg26$        & 0.485 (50.7\%) & 108.1 (41.1\%) & 104
(57.9\%) & 98 (23.9\%) & 204 (50.2\%) 

\enddata

\tablecomments{The table gives the spectral parameter $\gamma$, the length of the
skeleton (in radians), the number of minima, maxima, and saddle points as a
function of beam FWHM for the un-thresholded co-added \emph{WMAP} map
using the Kp0 mask. The numbers in parentheses are the percentage of
the 5000 simulations with a lower value than that found in the
\emph{WMAP} map. The results are for the full map, and for the
northern and southern hemispheres separately.}

\end{deluxetable*}

\begin{deluxetable*}{crrrrr}
\tablewidth{0pt} 
\tabletypesize{\small} 
\tablecaption{Power spectrum
dependent measurements, Kp0 mask using only $|b|>30^{\circ}$\label{tab:measurements_b30}}
\tablecolumns{6}
\tablehead{FWHM & $\gamma$\quad\quad\quad$\:$ & Length\quad\quad\quad &
$N_{\textrm{min}}$ \quad\quad$\:$& $N_{\textrm{max}}$ \quad\quad$\:$& $N_{\textrm{sad}}$ \quad\quad$\:$}
\startdata
%\cutinhead{Full sky}

\cutinhead{Northern hemisphere}
 $2\fdg13$	        & 0.538 (98.7\%) & 167.7 (57.5\%) & 327
(35.7\%) & 339 (75.5\%) & 664 (55.2\%) \\ 
 $2\fdg55$	        & 0.549 (99.1\%) & 134.9 (75.4\%) & 215
(34.3\%) & 219 (52.5\%) & 437 (52.0\%) \\ 
 $2\fdg98$	        & 0.561 (99.1\%) & 109.6 (38.7\%) & 152
(47.0\%) & 153 (53.3\%) & 311 (71.7\%) \\ 
 $3\fdg40$	        & 0.565 (98.1\%) & 92.2 (54.3\%) & 117
(82.7\%) & 106 (23.8\%) & 223 (60.0\%) \\ 
 $3\fdg83$	        & 0.566 (95.8\%) & 78.9 (69.6\%) & 89
(84.1\%) & 86 (70.4\%) & 173 (80.5\%) \\ 
 $4\fdg26$	        & 0.569 (92.9\%) & 69.4 (92.3\%) & 74
(97.3\%) & 67 (71.4\%) & 140 (94.3\%) \\ 

\cutinhead{Southern hemisphere}
 $2\fdg13$	        & 0.471 (38.5\%) & 166.1 (52.7\%) & 328
(48.9\%) & 328 (48.0\%) & 651 (38.3\%) \\ 
 $2\fdg55$	        & 0.476 (40.8\%) & 134.9 (83.2\%) & 227
(84.7\%) & 225 (78.5\%) & 453 (89.8\%) \\ 
 $2\fdg98$	        & 0.480 (40.2\%) & 113.6 (98.2\%) & 168
(97.6\%) & 172 (99.3\%) & 336 (99.3\%) \\ 
 $3\fdg40$	        & 0.488 (42.7\%) & 95.0 (91.2\%) & 126
(98.2\%) & 117 (78.3\%) & 244 (97.2\%) \\ 
 $3\fdg83$	        & 0.498 (45.0\%) & 80.7 (78.7\%) & 87
(67.9\%) & 85 (53.5\%) & 180 (89.6\%) \\ 
 $4\fdg26$	        & 0.505 (45.4\%) & 68.9 (44.5\%) & 70
(79.8\%) & 63 (31.5\%) & 137 (78.4\%)  
\enddata

\tablecomments{Same as Table~\ref{tab:measurements}, but where all areas
with $|b|<30^{\circ}$ have been included in the mask, i.e., excluded from the analysis.}

\end{deluxetable*}

The algorithms described in \S \ref{sec:algorithms} result in a number
of by-products, and several of these may serve as useful consistency
checks. In particular, the validity of the assumed power spectrum may
be tested by e.g., counting the number of stationary points, or by
measuring the total length of the un-thresholded skeleton. This is the
topic of the next subsection, while the results from the actual
non-Gaussianity analysis are presented in the following subsection.

In the following, the results from the measurements are presented
as a function of the smoothing scale. One should therefore remember that
each quantity is a continuous function of that smoothing scale, and
the results at two different scales cannot be considered independent.

\subsection{Power spectrum consistency statistics}

In Table \ref{tab:measurements}, the spectral index $\gamma$,
the skeleton length, and the stationary point count results for the
un-thresholded map using Kp0 as the base mask are shown, for both the full sky and for
the northern and southern hemispheres separately. The numbers in
parentheses show the percentage of simulations with a lower value than
that found in the \emph{WMAP} sky. The value 100\% is reserved for the
special case in which \emph{all} 5000 simulations have a lower
value. Note also that the percentages refer directly to the cumulative
distribution functions and not to the significance level as such;
both 2.5\% and 97.5\% indicate a $2\,\sigma$ effect.

On the smallest scales ($0\fdg53$ and $0\fdg64$ FWHMs) the
spectral parameter $\gamma$ is high, but well within the acceptable
range. This is also the case for the total length of the
skeleton. However, the numbers of stationary points are clearly not
acceptable -- only two simulations out of 5000 realizations have as
many saddle points as the \emph{WMAP} map at $0\fdg53$ FWHM, and the
overall stationary point counts lie at around $3\,\sigma$ compared with
the distribution of the 5000 simulated results. However, this is not a
very unsettling result, considering that our choice of base mask does
not discard the 700 known point sources. It should be
expected that point sources cause deviations on these scales. In fact,
these measurements are only intended to set a meaningful lower limit
on the smoothing FWHM to use in the following analyses.

At the other extreme, there are only a few hundred stationary points
in the full sky at the $4\fdg26$ scale.  Since the method for
computing the genus is based on counting stationary points, we must
expect artifacts at larger scales; therefore, we choose not to
extend our analysis to this regime.

On scales larger than $1\degr$, all full-sky numbers are in fairly
good agreement with the model, although the spectral parameter
$\gamma$ is somewhat large ($>93$\% of the simulations). However, the
situation is much more interesting when looking at the northern
hemisphere separately. In this case the \emph{WMAP} spectral parameter
$\gamma$ is higher than 98.9\% of the simulation on all scales,
reaching the value of 0.548 at $3\fdg40$, which is high at the 99.6\%
level. Except for the spectral parameter, all measurements are
perfectly acceptable. No sign of discrepancy is found in the southern
hemisphere.

In Table \ref{tab:measurements_b30}, we show the same statistics for a
number of smoothing scales in the northern and southern hemispheres
when all areas with $\vert b\vert <30\degr$ are added to the base
mask. Extending the excluded region only alters the results by small amounts in the
northern hemisphere, but in the southern hemisphere the stationary
point counts are relatively high for FWHMs around $3\degr$.

\subsection{The Minkowski functionals and the skeleton length}
\label{sec:nongauss_res}

The three Minkowski functionals and the length of the skeleton,
computed from maps smoothed with a $1\fdg28$ Gaussian beam (Kp0 base
mask), are shown in Figure \ref{fig:mink_full}. The \emph{WMAP} data
are plotted with black filled circles, while the median computed from
the ensemble of 5000 Monte Carlo simulations is shown as a thin solid
line. The 1, 2, and $3\,\sigma$ confidence bands around the median are
shown as gray bands. However, in all cases except the genus, the
confidence regions are so narrow that it is virtually impossible to
distinguish between the different elements.

\begin{figure*}
\plotone{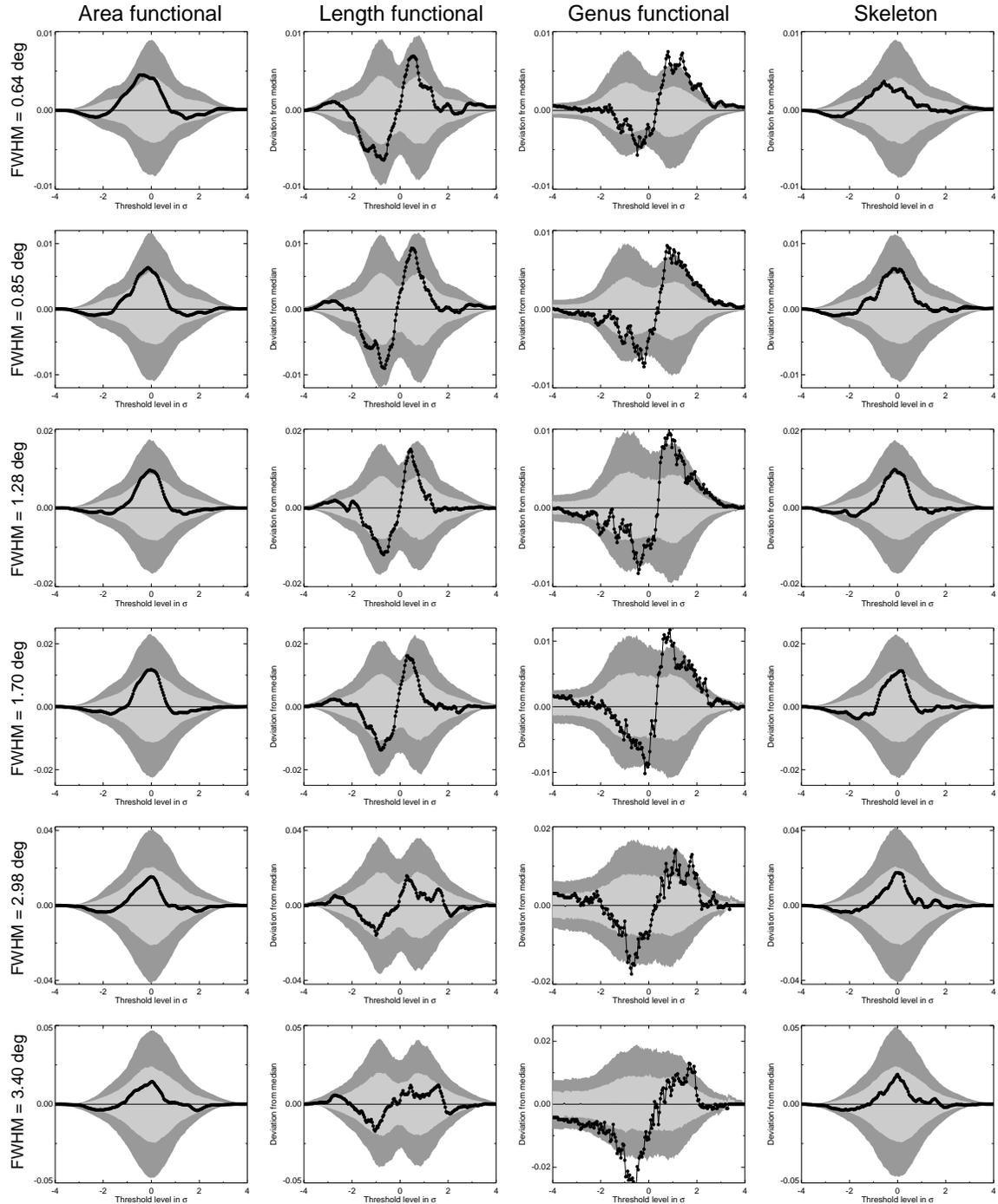}
\caption{Results from the Minkowski functional and skeleton
  measurements, with the median subtracted from each bin. Without such
  a subtraction it is not possible to distinguish between a function
  that follows the median and one that lies in the $2\,\sigma$ region,
  as a result of the extremely narrow confidence bands -- usually only a few
  percent wide. The gray bands indicate 1 and $2\,\sigma$ confidence
  regions as computed from the simulations.}
\label{fig:mink_funcs}
\end{figure*}

For this reason, we subtract in what follows the median from each bin
before plotting the functions, and such results are shown for the
full-sky measurements in Figure \ref{fig:mink_funcs}. In these plots
there are several points worth noticing: First, we see that the
genus confidence bands are asymmetric in the tails. This is because of
the inability of our method to deal with masks; i.e., on a cut sky,
the number of maxima plus the number of minima minus the number of
saddle-points is likely to be non-zero, and this translates in our
method into a non-zero genus at $\tilde\nu = -\infty$. On the other
hand, there are no stationary points at very high thresholds (when the
remaining area of the sky is very small); therefore, the genus will
always be zero here. However, this asymmetry is only a very minor
problem for thresholds between $-3\,\sigma_0$ and $3\,\sigma_0$, and the
effect is in any case calibrated by simulations.

Second, the plots show that the properties of the full-sky field on
scales up to $3\degr$ are in fairly good agreement with the model. No
single function shows excessive deviations, but the deviations are
also not too small compared to the confidence bands, as could easily
be the case if the power spectrum was flawed. However, the genus curve
does seem to have a somewhat large amplitude, and more so at large
scales. At $3\fdg40$ FWHM the genus lies clearly outside the
$2\,\sigma$ band.

In order to quantify the degree of deviation in each case, we use the
diagonal $\chi^2$ statistic described in \S \ref{sec:quantify}.  The
results from these measurements are shown in Table \ref{tab:chisq}
(for the Kp0 base mask) and Table \ref{tab:chisq_b30} (for the larger
base mask). The numbers in these tables indicate the percentage of
simulations with a \emph{lower} $\chi^2$ value than that of the
observed \emph{WMAP} data. Once again we see that the full-sky numbers
are in excellent agreement with the Gaussian theoretical model, except
for the genus at large ($3\degr$--$4\degr$) scales, which has a $\chi^2$
that is high at the 96\% level.

\begin{deluxetable}{ccccc}
\tablewidth{0pt}
\tabletypesize{\small}
\tablecaption{Minkowski functional and length of skeleton $\chi^2$
results, Kp0 mask\label{tab:chisq}}  
\tablecolumns{5}
\tablehead{FWHM & Area & Length & Genus & Skeleton}
\startdata

\cutinhead{Full sky}
 $0\fdg53$  		& 49.4\% & 67.8\% & 86.2\% & 23.4\%   \\ 
 $0\fdg64$ 		& 43.9\% & 65.2\% & 88.3\% & 28.6\%   \\ 
 $0\fdg85$ 		& 35.2\% & 56.8\% & 89.4\% & 35.4\%   \\ 
 $1\fdg28$ 		& 26.9\% & 44.3\% & 89.3\% & 29.3\%   \\ 
 $1\fdg70$ 		& 23.5\% & 34.8\% & 86.0\% & 28.4\%   \\ 
 $2\fdg13$ 		& 21.5\% & 29.1\% & 79.5\% & 24.8\%   \\ 
 $2\fdg55$ 		& 19.9\% & 19.1\% & 77.5\% & 20.8\%   \\ 
 $2\fdg98$ 		& 16.2\% & 13.8\% & 69.3\% & 16.6\%   \\ 
 $3\fdg40$ 		& 12.5\% & 10.8\% & 95.9\% & 11.3\%   \\ 
 $3\fdg83$ 		& \phn8.2\% & \phn7.6\% & 94.3\% & \phn8.1\%   \\ 
 $4\fdg26$\tablenotemark{*}   & \phn4.3\% & \phn3.1\% & 80.5\% & \phn7.8\%   \\ 

\cutinhead{Northern hemisphere}
 $0\fdg85$ 		& 12.5\% & 34.5\% & 95.9\% & 14.4\%   \\ 
 $1\fdg28$ 		& 15.4\% & 32.5\% & 95.3\% & 15.1\%   \\ 
 $1\fdg70$ 		& 14.1\% & 24.6\% & 97.2\% & 18.9\%   \\ 
 $2\fdg13$ 		& \phn9.6\% & 26.7\% & 95.7\% & 13.8\%   \\ 
 $2\fdg55$ 		& \phn9.4\% & 30.6\% & 98.4\% & 18.4\%   \\ 
 $2\fdg98$ 		& 13.1\% & 34.6\% & 98.4\% & 23.0\%   \\ 
 $3\fdg40$ 		& 17.7\% & 35.3\% & 99.5\% & 20.3\%   \\ 
 $3\fdg83$ 		& 20.3\% & 29.6\% & 94.9\% & 32.2\%   \\ 
 $4\fdg26$\tablenotemark{*} & 28.3\% & 29.4\% & 77.3\% & 40.1\%   \\

\cutinhead{Southern hemisphere}
 $0\fdg85$ 		& 33.5\% & 54.9\% & 27.7\% & 33.6\%   \\ 
 $1\fdg28$ 		& 53.2\% & 70.9\% & 37.7\% & 58.4\%   \\ 
 $1\fdg70$ 		& 59.2\% & 64.0\% & 43.0\% & 67.0\%   \\ 
 $2\fdg13$ 		& 58.6\% & 65.7\% & 28.8\% & 55.9\%   \\ 
 $2\fdg55$ 		& 58.6\% & 71.4\% & 19.4\% & 54.6\%   \\ 
 $2\fdg98$ 		& 54.2\% & 70.5\% & 85.2\% & 49.5\%   \\ 
 $3\fdg40$ 		& 51.4\% & 66.6\% & 57.6\% & 43.9\%   \\ 
 $3\fdg83$ 		& 45.0\% & 56.3\% & 21.8\% & 35.5\%   \\ 
 $4\fdg26$\tablenotemark{*}   & 39.6\% & 40.4\% & \phn9.6\% & 32.2\%   

\enddata

\tablecomments{Area functional, length functional, and genus results
from diagonal $\chi^2$ tests, as computed from values of the threshold
$\tilde\nu$ between $-3\,\sigma_0$ and $3\,\sigma_0$. The numbers indicate
the percentage of simulated realizations with a $\chi^2$ value lower
than that for the co-added \emph{WMAP} map using the Kp0 mask.}

\tablenotetext{*}{$\chi^2$ computed only for thresholds between
  $-2.5\,\sigma_0$ and $2.5\,\sigma_0$.}

\end{deluxetable}

\begin{deluxetable}{ccccc}
\tablewidth{0pt}
\tabletypesize{\small}
\tablecaption{Minkowski functional and skeleton $\chi^2$ results, Kp0
mask using only $|b|>30^{\circ}$\label{tab:chisq_b30}} 
\tablecolumns{5}
\tablehead{FWHM & Area & Length & Genus & Skeleton}
\startdata

\cutinhead{Northern hemisphere}
 $2\fdg13$ 		& 22.5\% & 48.0\% & 97.6\% & 22.6\%   \\ 
 $2\fdg55$ 		& 11.3\% & 49.8\% & 99.2\% & 20.2\%   \\ 
 $2\fdg98$ 		& \phn9.2\% & 49.8\% & 99.9\% & 33.2\%   \\ 
 $3\fdg40$ 		& 10.6\% & 46.0\% & 94.5\% & 29.5\%   \\ 
 $3\fdg83$ 		& 12.3\% & 40.2\% & 62.2\% & 24.5\%   \\ 
 $4\fdg26$\tablenotemark{*} 	        & 12.1\% & 27.5\% & 42.0\% & 16.5\%   \\ 

\cutinhead{Southern hemisphere}
 $2\fdg13$ 		& 31.5\% & 18.7\% & 16.6\% & 35.9\%   \\ 
 $2\fdg55$ 		& 36.2\% & 12.7\% & \phn1.0\% & 36.8\%   \\ 
 $2\fdg98$ 		& 31.5\% & \phn7.2\% & 34.1\% & 36.3\%   \\ 
 $3\fdg40$ 		& 34.6\% & 13.6\% & 17.1\% & 25.0\%   \\ 
 $3\fdg83$ 		& 33.5\% & 16.2\% & 96.8\% & 19.2\%   \\ 
 $4\fdg26$\tablenotemark{*} 	        & 32.0\% & 15.5\% & 78.9\% & \phn6.1\%   

\enddata
\tablecomments{Same as Table~\ref{tab:chisq}, but where all areas
with $|b|<30^{\circ}$ have been included in the mask.}
\tablenotetext{*}{$\chi^2$ computed only for thresholds between
  $-2.5\,\sigma_0$ and $2.5\,\sigma_0$.}

\end{deluxetable}

This pattern is clearly stronger in the northern hemisphere. Here the
genus $\chi^2$ generally is very large, with a pronounced peak at
$3\fdg40$. At that scale only 0.5\% of the simulations have a larger
$\chi^2$. The other functionals are all perfectly acceptable.  In the
southern hemisphere, even the genus is acceptable, and no signs of
discrepancy between the observed sky and the simulations are found
there. We see that these results are not particularly dependent on the
choice of base mask, except that with the larger mask, the genus
$\chi^2$ peaks at $2\fdg98$ FWHM in the northern hemisphere, rather
than at $3\fdg40$, and that it is curiously low at $2\fdg55$ in the
southern hemisphere (at the 1\% level). However, it is difficult to
attach much significance to the latter result, since it is highly
unstable with respect to both Galactic cut and smoothing scale.

\begin{figure*}
\plotone{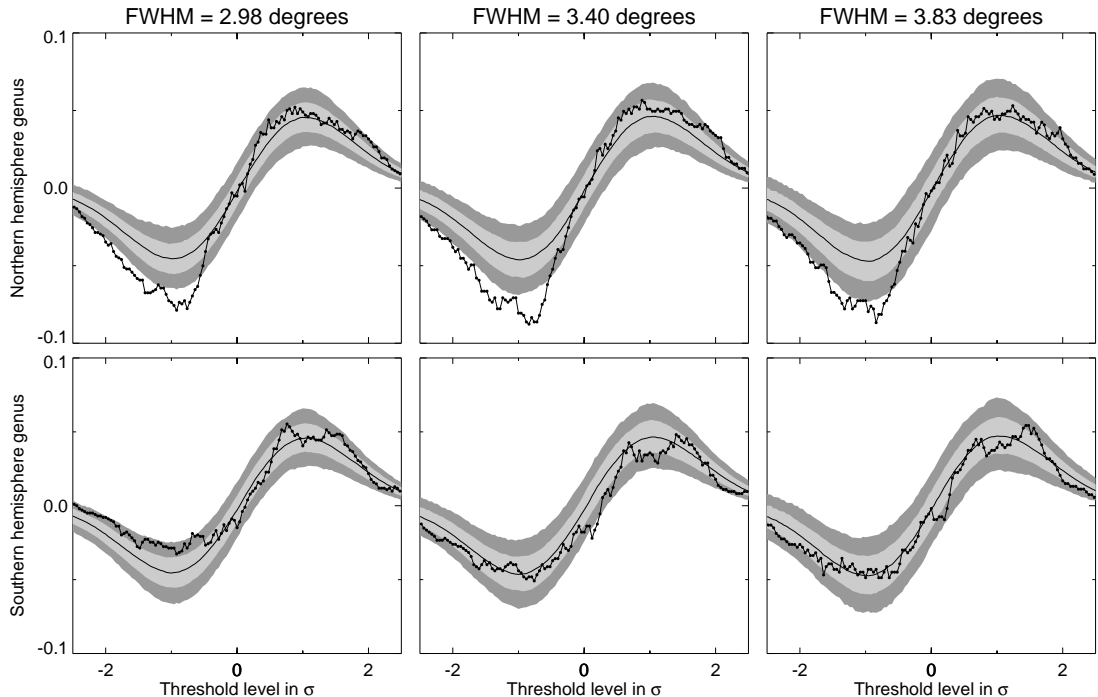}
\caption{Large-scale genus measurements. The northern hemisphere genus
functional at $3\fdg40$ FWHM has a $\chi^2$ value that is larger
than that found in 99.5\% of the simulations, and only 1 out of 5000
simulations has a larger negative threshold amplitude. Conversely,
the southern hemisphere genus functional at $2\fdg98$ FWHM has
a very small amplitude, and its asymmetry parameter is extreme at the
99\% level.}
\label{fig:genus_largescale}
\end{figure*}

The $\chi^2$ statistic only measures the overall level of deviation
between a given function and the theoretical mean. It is possible that
other statistics could be more sensitive to particular features. To
study the behavior of the genus in more detail, we therefore fit the
parametric function given by Equation (\ref{eq:genus_par}) to each
realization \citep{Park:2004}, allowing for non-zero shifts and
arbitrary amplitudes. The results from these computations are shown in
Tables \ref{tab:genus} (Kp0 base mask) and \ref{tab:genus_b30} (larger
base mask). Only the amplitudes and the asymmetry parameters are
shown; the shift parameters are in all cases well inside the $2\,\sigma$
range, except for the smallest scales ($0\fdg53$ and $0\fdg64$ FWHM)
. The number in parentheses here shows the percentage of the simulated
maps having the fitted parameter \emph{smaller} than for the
\emph{WMAP} data.

We see that the genus amplitude $A$ (and especially when fitted for
$\tilde\nu$ in the range $-2.5\,\sigma_0$ to $-0.2\,\sigma_0$, i.e.,
$A_{-}$) is very high in the northern hemisphere and relatively small
in the southern hemisphere, compared with the simulations.  In
particular, the $3\fdg40$ scale once again stands out as special
($2\fdg98$ scale with the $|b|>30^{\circ}$ base mask); in the
northern hemisphere the negative threshold amplitude is so large that
only 1 out of 5000 simulations has a larger amplitude. Using the
larger base mask, \emph{no} simulations have as large $A_{-}$ as the
\emph{WMAP} map at $2\fdg98$ FWHM. In adition, the asymmetry parameter
$\Delta g$ is clearly large at the $2\,\sigma$ level in the northern
hemisphere.

\begin{deluxetable*}{crrrr}
\tablewidth{0pt}
\tabletypesize{\small}
\tablecaption{Genus results, Kp0 mask\label{tab:genus}} 
\tablecolumns{5}
\tablehead{FWHM& $A$\quad\quad\quad\quad& $A_{-}$$\,\,\,$\quad\quad\quad&
  $A_{+}$\quad\quad\quad& $\Delta g$\quad\quad\quad} 
\startdata

\cutinhead{Full sky}
 $0\fdg53$	       	& 0.087 (87.3\%) & 0.086 (82.6\%) & 0.091
(98.0\%) & -0.024 (\phn5.5\%)  \\ 
 $0\fdg64$       	& 0.092 (87.4\%) & 0.090 (79.3\%) & 0.096
(98.9\%) & -0.034 (\phn2.3\%)  \\ 
 $0\fdg85$	       	        & 0.091 (91.5\%) & 0.089 (86.4\%) & 
0.094 (98.3\%) & -0.027 (\phn8.1\%)  \\ 
 $1\fdg28$	       	        & 0.082 (91.9\%) & 0.080 (85.4\%) & 
0.087 (98.4\%) & -0.038 (10.3\%)  \\ 
 $1\fdg70$	       	        & 0.078 (90.4\%) & 0.076 (84.2\%) & 
0.084 (98.6\%) & -0.049 (11.3\%)  \\ 
$2\fdg13$	       	        & 0.080 (95.3\%) & 0.080 (93.8\%) & 
0.080 (94.4\%) & 0.000 (53.9\%)  \\ 
$2\fdg55$	       	        & 0.082 (96.4\%) & 0.083 (96.0\%) & 
0.080 (92.2\%) & 0.018 (67.5\%)  \\ 
 $2\fdg98$	       	        & 0.081 (93.6\%) & 0.083 (92.9\%) & 
0.083 (93.8\%) & -0.001 (55.2\%)  \\ 
 $3\fdg40$	       	        & 0.095 (99.5\%) & 0.101 (99.6\%) & 
0.081 (86.8\%) & 0.111 (93.5\%)  \\ 
 $3\fdg83$	       	        & 0.096 (99.1\%) & 0.104 (99.5\%) & 
0.081 (82.0\%) & 0.128 (93.2\%)  \\ 
$4\fdg26$	               & 0.094 (96.9\%) & 0.097 (96.1\%) & 
0.091 (93.9\%) & 0.031 (67.7\%)  \\ 

\cutinhead{Northern hemisphere}
 $0\fdg85$          & 0.100 (97.5\%) & 0.099 (96.9\%) &
 0.104 (99.7\%) & -0.021 (18.3\%)  \\
 $1\fdg28$          & 0.090 (93.9\%) & 0.089 (91.0\%) &
0.094 (98.4\%) & -0.027 (20.4\%)  \\
 $1\fdg70$          & 0.091 (97.9\%) & 0.091 (96.6\%) &
0.092 (97.7\%) & -0.004 (47.3\%)  \\
 $2\fdg13$          & 0.095 (98.7\%) & 0.097 (98.8\%) &
0.087 (89.4\%) & 0.056 (85.4\%)  \\
 $2\fdg55$          & 0.104 (99.9\%) & 0.110 (99.9\%) &
0.085 (81.0\%) & 0.132 (97.6\%)  \\
 $2\fdg98$          & 0.110 (99.9\%) & 0.120 (99.9\%) &
0.085 (76.8\%) & 0.170 (98.3\%)  \\
 $3\fdg40$          & 0.122 (99.9\%) & 0.134 (99.9\%) &
0.092 (85.7\%) & 0.188 (97.5\%)  \\
 $3\fdg83$          & 0.116 (99.5\%) & 0.129 (99.7\%) &
0.086 (65.8\%) & 0.204 (97.0\%)  \\
 $4\fdg26$         & 0.113 (97.7\%) & 0.123 (98.0\%) &
 0.094 (77.3\%) & 0.136 (87.2\%)  \\

\cutinhead{Southern hemisphere}
 $0\fdg85$          & 0.084 (23.3\%) & 0.081 (15.5\%) &
 0.088 (46.3\%) & -0.039 (\phn5.6\%)  \\
 $1\fdg28$          & 0.077 (38.6\%) & 0.074 (26.6\%) &
 0.081 (58.9\%) & -0.044 (10.7\%)  \\
 $1\fdg70$          & 0.070 (26.1\%) & 0.066 (15.9\%) &
 0.078 (60.7\%) & -0.083 (\phn4.4\%)  \\
 $2\fdg13$          & 0.069 (28.7\%) & 0.065 (19.4\%) &
 0.076 (54.0\%) & -0.076 (12.4\%)  \\
 $2\fdg55$          & 0.068 (27.0\%) & 0.062 (16.8\%) &
0.077 (57.9\%) & -0.108 (\phn9.2\%)  \\
 $2\fdg98$          & 0.058 (\phn7.6\%) & 0.047
 (\phn2.4\%) & 0.083 (69.9\%) & -0.271 (\phn0.4\%)  \\
 $2\fdg98$\tablenotemark{*}          & 0.064 (16.0\%) & 
0.054 (\phn6.5\%) & 0.094 (89.2\%) & -0.269 (\phn0.9\%)  \\
 $3\fdg40$          & 0.078 (54.2\%) & 0.082 (63.3\%) &
0.067 (20.3\%) & 0.100 (86.0\%)  \\
 $3\fdg83$          & 0.082 (58.6\%) & 0.083 (60.7\%) &
0.077 (43.2\%) & 0.036 (65.4\%)  \\
 $4\fdg26$         & 0.086 (63.2\%) & 0.083 (55.4\%) & 0.089
  (67.8\%) & -0.038 (43.3\%)  

\enddata 
\tablecomments{The fitted amplitude of the genus for the full range in
$\tilde\nu$, for negative $\tilde\nu$, and for positive $\tilde\nu$,
and the genus asymmetry parameter, using the co-added \emph{WMAP} map
with Kp0 mask. The numbers in parentheses show the percentage of the
5000 simulations with a value smaller than for the \emph{WMAP} map.}

\tablenotetext{*}{The two large cold spots discussed by \citet{Park:2004}
  are masked out.}

\end{deluxetable*}

\begin{deluxetable*}{rrrrr}
\tablewidth{0pt}
\tabletypesize{\small}
\tablecaption{Genus results, Kp0 mask, using only $|b|>30^{\circ}$\label{tab:genus_b30}} 
\tablecolumns{5}
\tablehead{FWHM& $A$\quad\quad\quad\quad& $A_{-}$$\,\,\,$\quad\quad\quad&
$A_{+}$\quad\quad\quad& $\Delta g$\quad\quad\quad} 
\startdata

\cutinhead{Northern hemisphere}
 $2\fdg13$ 		& 0.106 (99.4\%) & 0.110 (99.4\%) &
0.094 (88.8\%) & 0.079 (88.6\%)  \\  
 $2\fdg55$ 		& 0.115 (99.8\%) & 0.124 (99.9\%) &
0.091 (76.7\%) & 0.155 (97.0\%)  \\  
 $2\fdg98$ 		& 0.131 (99.9\%) & 0.144 (\phd100\%) &
0.095 (79.8\%) & 0.206 (97.8\%)  \\  
 $3\fdg40$ 		& 0.125 (99.5\%) & 0.137 (99.6\%) &
0.101 (83.0\%) & 0.151 (90.8\%)  \\  
 $3\fdg83$ 		& 0.117 (95.4\%) & 0.129 (96.2\%) &
0.099 (73.2\%) & 0.132 (83.5\%)  \\  
 $4\fdg26$         & 0.120 (93.4\%) & 0.131 (92.8\%) &
0.106 (77.2\%) & 0.107 (75.6\%)  \\  

\cutinhead{Southern hemisphere}
 $2\fdg13$ 		& 0.078 (43.7\%) & 0.077 (40.9\%) &
0.084 (64.0\%) & -0.044 (27.7\%)  \\  
 $2\fdg55$ 		& 0.079 (46.2\%) & 0.079 (44.5\%) &
0.081 (48.5\%) & -0.014 (45.7\%)  \\  
 $2\fdg98$ 		& 0.068 (15.3\%) & 0.064 (14.1\%) &
 0.077 (32.8\%) & -0.088 (21.2\%)  \\  
 $3\fdg40$ 		& 0.077 (33.6\%) & 0.082 (45.6\%) &
0.065 (10.3\%) & 0.117 (85.6\%)  \\  
 $3\fdg83$ 		& 0.108 (89.1\%) & 0.127 (96.6\%) &
0.064 (\phn8.4\%) & 0.334 (99.6\%)  \\  
 $4\fdg26$ 	& 0.104 (79.1\%) & 0.123 (90.8\%) &
0.069 (13.0\%) & 0.279 (97.5\%)  

\enddata 

\tablecomments{Same as Table \ref{tab:genus}, but with all areas with
$|b|< 30^{\circ}$ added to the mask.}

\end{deluxetable*}

In the southern hemisphere, the genus amplitudes are relatively, but
not extremely, low. However, here the asymmetry parameter is rather
small, at the 99\% level, for $2\fdg98$ smoothing and use of the Kp0
base mask. This result is in fact quite peculiar, and it is difficult
to decide how much significance one should attach to it; it is very
sensitive to smoothing scale, and it is not seen in the larger base
mask measurement. One could suspect that this result is caused by some
features near the Galactic plane. However, we have not been able to
locate any region that by exclusion brings the results to an
acceptable level. One example from this search is shown in Table
\ref{tab:genus}, marked by an asterisk.  In this case we have removed two
disks of $30^{\circ}$ radius centered on $(l,b) = (330^{\circ},
-10^{\circ})$ and $(200^{\circ}, -20^{\circ})$, corresponding to the
two large cold spots discussed by \citet{Park:2004}. Obviously, the
peculiar 99\% result is not connected to these regions.

\begin{figure}
\mbox{\epsfig{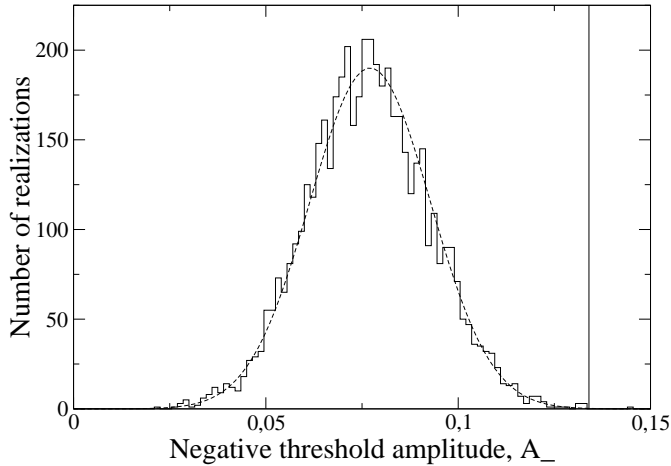}}
%\plotone{f6.eps}
\caption{Results from measurements of $A_{-}^{\textrm{north}}$ at
  $3\fdg40$ FWHM. The best-fit Gaussian distribution has a mean of
  0.0771 and a standard deviation of 0.0157; therefore, the
  observed \emph{WMAP} value of 0.134 (\emph{thin vertical line}) formally
  corresponds to a $3.6\,\sigma$ effect.}
\label{fig:aneg}
\end{figure}

The genus for both the northern and the southern hemispheres at
$2\fdg98$, $3\fdg40$, and $3\fdg83$ FWHMs is shown in Figure
\ref{fig:genus_largescale}. Here we clearly see the origin of the
extreme numbers found above: the northern hemisphere genus at around
$3\fdg40$ FWHM lies well outside the lower $2\,\sigma$ confidence
region and even outside the $3\,\sigma$ region for an extended range of
thresholds. In addition, the genus has a very small amplitude at negative
thresholds on the southern hemisphere, but an average amplitude at
positive thresholds, resulting in a rather extreme asymmetry
parameter.

The distribution of negative threshold amplitudes $A_{-}$ for the
northern hemisphere at $3\fdg40$ FWHM is shown for the 5000
simulations in Figure \ref{fig:aneg}, together with the observed
\emph{WMAP} value of 0.134 (\emph{vertical line}). The distribution is
close to Gaussian with a mean of 0.0771 and a standard deviation of
0.0157. Thus, the observed \emph{WMAP} $A_{-}$-value is formally
larger than the simulations at $3.6\,\sigma$. However, this number is
likely to be slightly overestimated: since the amplitude is a strictly
positive number, we should expect the true distribution to be somewhat
skewed toward large values. On the other hand, this effect cannot be
very significant in our case, given the good fit between the histogram
and the Gaussian approximation. The case seems solid for rejecting the
null hypothesis at considerably more than $3\,\sigma$ confidence.

\subsection{Comparison with alternative maps}

So far we have only studied the co-added \emph{WMAP} map, which is
dominated by the $Q$-band. The influence of residual foregrounds could
therefore be an important concern when interpreting these results. In
order to study this issue closer, we now reestimate the northern
hemisphere statistics (using the Kp0 mask), this time for a set of
seven different maps: (1) the co-added \emph{WMAP} map, (2) the
Tegmark et al.\ cleaned map (Tegmark, de Oliveira-Costa, \& Hamilton
2003), (3) the \emph{WMAP} Internal Linear Combination (ILC) map
\citep{Bennett:2003b}, (4)--(6) the averaged $Q$-, $V$- and $W$-bands,
and (7) a synchrotron-corrected map [defined by $[2.65 \:\textrm{Ka} -
\textrm{K}]/1.65$; see Vielva et al.\ 2004)].

Each of these maps has its own effective beam profile, and in order
for a direct comparison to be meaningful, we have to pre-smooth each
map so that they have a common resolution. This was achieved by first
deconvolving the old beam and subsequently convolving with a
$1^{\circ}$ Gaussian beam. The only exception is the ILC map, which is
already smoothed to the appropriate resolution.

The simulations that we carry out in this paper are very
CPU-intensive, and it would therefore be very time consuming to
perform a complete analysis of each of the above-mentioned
maps. However, after pre-smoothing each map to a common resolution,
only the effect of noise can potentially modify the results, and on
the large scales that we consider, this effect is very small. We may
therefore instead use the existing simulations when interpreting
the new results.

\begin{figure*}
\mbox{\epsfig{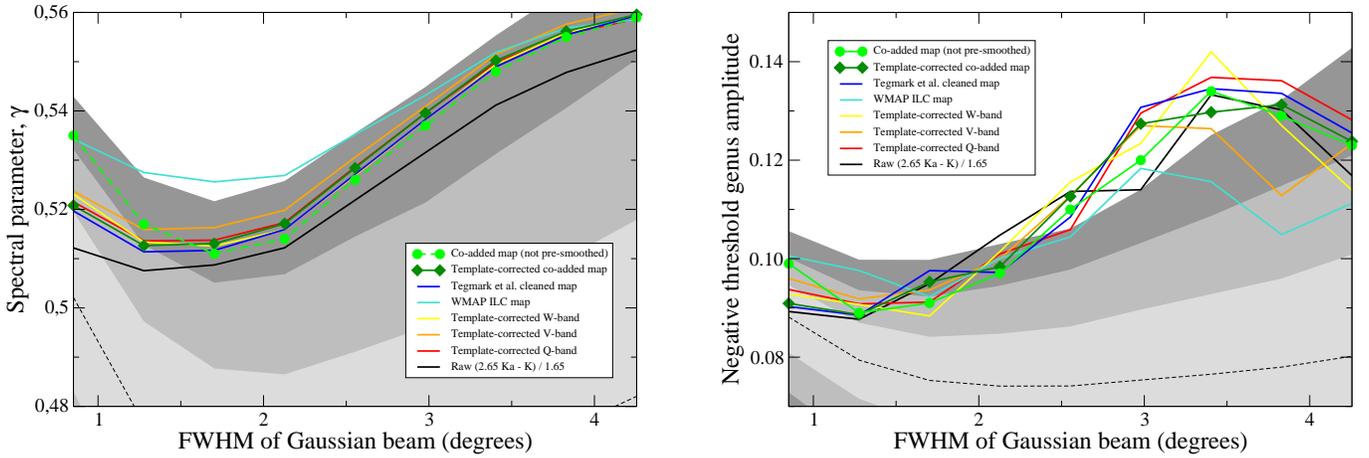}}
\caption{Scale dependence of the spectral parameter $\gamma$ (\emph{left}) and
  the negative threshold genus amplitude $A_{-}$ (\emph{right}),
  computed on the northern hemisphere for seven different maps. The
  gray bands show the 1, 2, and $3\,\sigma$ confidence regions
  computed from simulations. All maps have been pre-smoothed to a
  common $1^{\circ}$ resolution, except for the co-added \emph{WMAP}
  map, which is plotted both with and without pre-smoothing. The
  confidence bands are computed from simulations that are \emph{not}
  pre-smoothed.}
\label{fig:freq_dep}
\end{figure*}

The two panels of Figure \ref{fig:freq_dep} show the results from this
analysis for the spectral parameter $\gamma$ and the negative
threshold genus amplitude $A_{-}$. Each panel contains two sets of
fundamentally different elements. First, the gray 1, 2, and $3\,\sigma$
confidence regions and the light green curve with green filled circles (the
co-added \emph{WMAP} map) show the results of \S
\ref{sec:nongauss_res} and are thus not pre-smoothed to $1^{\circ}$
resolution. The seven remaining curves correspond to the maps listed
above and are pre-smoothed.

The effect of the pre-smoothing operation may be reconstructed by
comparing the two green curves (\emph{with circles and with diamonds}), which
show the results for the co-added \emph{WMAP} map, un-smoothed and
smoothed, respectively. For the spectral parameter $\gamma$ (\emph{left
panel}), we see that the pre-smoothing process effectively suppresses
$\gamma$ on small scales, but increases it very slightly on large
scales. However, this effect is both small and stable on large scales,
and is thus not difficult to account for. The effect on the genus
amplitude (\emph{right panel}) is slightly less predictable but follows the
same pattern; the pre-smoothed function lies generally a little higher
than the un-smoothed function on large scales.

\begin{figure*}
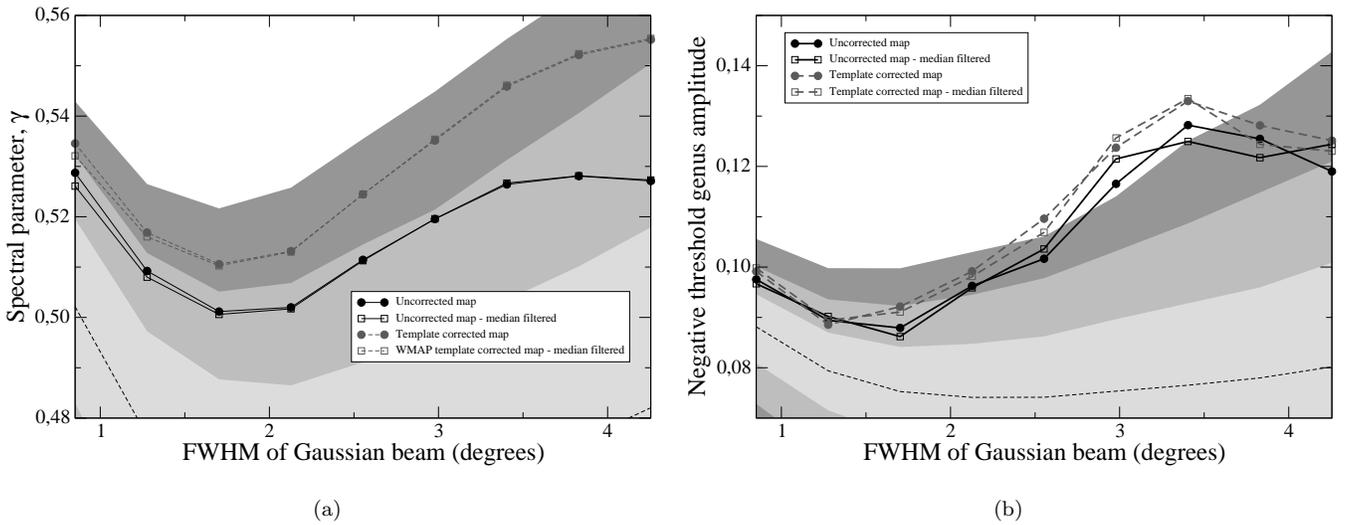

\mbox{\subfigure[]{\epsfig{figure=f9a.eps,width=0.48\textwidth,clip=}}
      \quad
      \subfigure[]{\epsfig{figure=f9b.eps,width=0.48\textwidth,clip=}}
}
\caption{Effect of point sources on the spectral parameter
  $\gamma$ (\emph{left}) and the negative threshold genus amplitude
$A_{-}$ (\emph{right}). The
  functions are computed on the northern hemisphere, restricted to the
  Kp0 base mask. Circles indicate the functions computed from the raw
  maps, and squares indicate the functions computed from
  median-filtered maps.}
\label{fig:median_filter}
\end{figure*}

The main conclusion to be drawn from these plots, however, is that the
quantities discussed above show a remarkable stability with respect to
both frequency and foreground correction method. If we neglect the
synchrotron-corrected map and the \emph{WMAP} ILC map, we see that the
scatter in $\gamma$ is much less than $1\,\sigma$, as seen
by the width of the confidence regions. The same is true for the genus
amplitude on the scale of most interest, namely, $2\fdg98$ FWHM. We
note that the \emph{WMAP} team warns (see the LAMBDA Web site) that
the ILC map should not be used for CMB studies, and indeed it is the
most deviant with respect to the other maps in the analysis.

The final issue to consider is the effect of point sources. We
therefore now apply a median filter to the co-added map, compute the
same set of quantities for this filtered map, and compare the results
to those of the unfiltered map. Our median filter is implemented by
replacing each pixel value that is excluded by the \emph{WMAP}
point-source mask, i.e., the mask consisting of about 700 disks of
$0\fdg5$ radius, by the median evaluated over a $1\degr$ disk
around that pixel. Such filters are commonly used in, e.g., radio
astronomy, to eliminate outliers.

The results from this experiment are shown in Figure
\ref{fig:median_filter}. Once again, we plot $\gamma$ and $A_{-}$
estimated in the northern hemisphere (Kp0 mask), since these
quantities represent the two strongest detections presented in this
paper. The results are shown for two different maps, the raw co-added
map and the \emph{WMAP} template-corrected map. In both cases the
effect of the median filter may be seen by comparing the solid and
dashed lines (circles and squares), which correspond to the raw map
and the median filtered map, respectively.

It is apparent that median filtering changes $\gamma$ by a negligible
amount on large scales while on small scales $\gamma$ is very
slightly decreased. The effect on the genus amplitude $A_{-}$ is less
predictable, as the two curves lie both above and below each
other. However, it is interesting to note that the filtered maps are
associated with a slightly larger amplitude than the raw maps at
$2\fdg98$ FWHM and a slightly smaller amplitude at $3\fdg83$ FWHM. Of
course, it is difficult to estimate the significance of this. In any
case, it is clear that the impact of point sources is too small to
explain the reported large-scale results in this paper.

\subsection{Correlations between the measurements}

\begin{figure*}
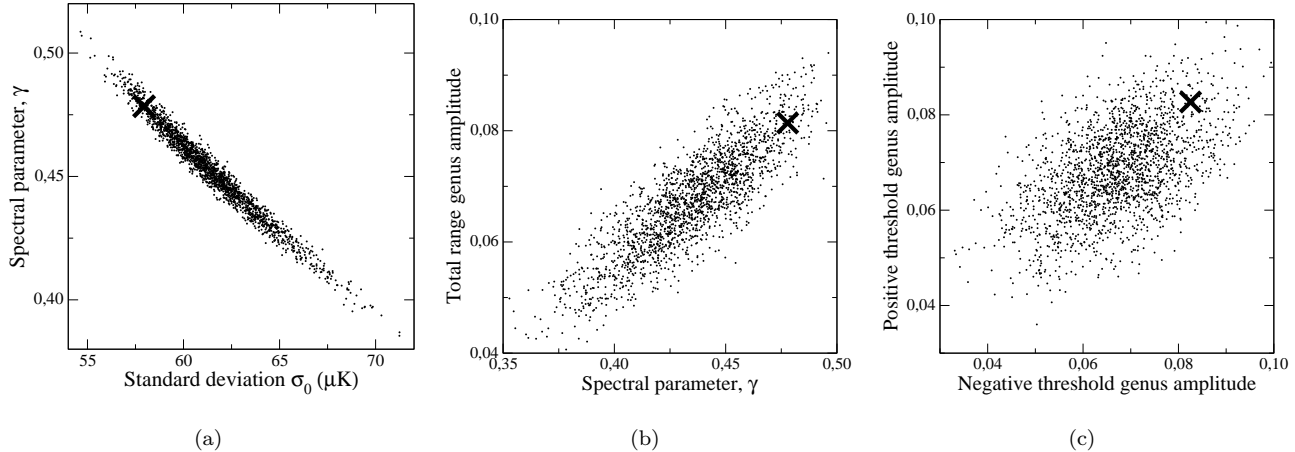

\mbox{\subfigure[]{\label{fig:correlations_a}\epsfig{figure=f10a.eps,width=0.3\textwidth,clip=}}
      \quad
      \subfigure[]{\label{fig:correlations_b}\epsfig{figure=f10b.eps,width=0.3\textwidth,clip=}}
      \quad
      \subfigure[]{\label{fig:correlations_c}\epsfig{figure=f10c.eps,width=0.3\textwidth,clip=}}}
\caption{Correlation plots of scalar quantity pairs, computed from
  maps smoothed with a $1\fdg28$ FWHM beam and constrained to the Kp0
  base mask. Each small filled circle corresponds to 1 of 2000 Monte
  Carlo realizations, and the cross marks the \emph{WMAP} value. The
  plotted relationships are (\emph{a}) the temperature standard
  deviation $\sigma_0$ versus the spectral parameter $\gamma$,
  (\emph{b}) the spectral parameter $\gamma$ versus the full range
  genus amplitude $A$, and (\emph{c}) the negative threshold genus
  amplitude $A_{-}$ versus the positive threshold amplitude $A_{+}$.}
\label{fig:correlations}
\end{figure*}

In the previous sections we presented measurements of several
important statistics. In this section we go one step further and seek
to understand the correlations between some of these statistics. This
goal is important both in order to understand the nature of the
quantities themselves and for assessing the combined significance of
the quoted results. In the following we focus on the full-sky
measurements smoothed with a $1\fdg28$ FWHM Gaussian beam.

Figure \ref{fig:correlations} shows the following scalar quantities
plotted pair-wise against each other: the temperature standard
deviation $\sigma_0$ versus the spectral parameter $\gamma$, the
spectral parameter $\gamma$ versus the full range genus amplitude $A$,
and the negative threshold genus amplitude $A_{-}$ versus the positive
threshold amplitude $A_{+}$. The values from each simulated
realization are marked by a small filled circle, and the \emph{WMAP}
results are marked by a cross.

In Figure \ref{fig:correlations}\emph{a} we see that the correlation between
$\sigma_0$ and $\gamma$ is very strong and negative. Of course, given
that $\gamma \equiv \sigma_1^2/(\sigma_0 \sigma_2)$, this is no
surprise. Nevertheless, it is well worth noticing that the spectral
parameter $\gamma$ may in many respects be identified with the
standard deviation of the map.

Figure \ref{fig:correlations}\emph{b} demonstrates the clear
correlation between the spectral parameter $\gamma$ and the full range
genus amplitude $A$. Although the scatter is non-negligible, a high
$\gamma$-value is probably accompanied by a large genus
amplitude. Thus, given the large $\gamma$-value of \emph{WMAP}, the
large observed genus amplitude seen in both Table \ref{tab:genus} and
Figure \ref{fig:mink_funcs} is not unexpected.

Finally, in Figure \ref{fig:correlations}\emph{c} we see that there is only a
very weak correlation between the negative threshold genus amplitude
$A_{-}$ and the positive threshold amplitude $A_{+}$.

The conclusions to be drawn from these plots are therefore the following:
\begin{itemize}
\item The spectral parameter $\gamma$ is almost inversely
  proportional to the standard deviation of the map.
\item The spectral parameter $\gamma$ and the genus amplitudes are
  correlated, and the corresponding numbers in Tables
  \ref{tab:measurements} and Table \ref{tab:genus} may not be
  considered as independent results.
\item The negative and positive threshold genus amplitudes are only
  very mildly correlated.
\end{itemize}

Given these relations, we may now make a few connections between the
results presented in \S \ref{sec:nongauss_res}. First, we noted that
the \emph{WMAP} spectral parameter $\gamma$ was large in both the
full-sky measurements and particularly the northern
hemisphere. However, it was not significantly different from the
simulations in the southern hemisphere. These results are thus in
perfect agreement with the genus measurements shown in Figures
\ref{fig:mink_funcs} and \ref{fig:genus_largescale}, and the
parameters given in Table \ref{tab:genus} --- both $\gamma$ and the
genus amplitudes are large in the northern hemisphere, but fully
acceptable in the southern hemisphere. The measurements of $\gamma$
and the genus amplitudes may obviously not be regarded as independent.

One further interesting connection may be made through the relation
between $\gamma$ and the standard deviation $\sigma_0$, which may be
taken as a very crude measure of the overall power in the map,
determined primarily by the large-scale modes. Since $\gamma$ is very
large in the northern hemisphere, the standard deviation is very
small; in other words, there is little power in this region. This
result therefore supports the conclusions of \citet{Eriksen:2004},
that there is a clear lack of large-scale power in the northern
hemisphere. Conversely, the genus results are generally stronger than
the $\gamma$ results, and given the relatively high $\Delta g$
results, a complete explanation must probably be formulated as a
combination of power spectrum and other non-Gaussianity effects.

\section{Conclusions}
\label{sec:conclusions}

In this paper we have presented the three two-dimensional Minkowski
functionals and the length of the skeleton of the co-added \emph{WMAP}
map, as well as a number of other statistics, and compared these with
simulations based on Gaussian perturbations with a best-fit power
spectrum and \emph{WMAP}-specific noise and beam properties. On scales
between $1\degr$ and $4\degr$, all the statistics except for the genus
Minkowski functional accept the model. On scales smaller than $\sim
0\fdg9$ there are too many stationary points, a result that
is interpreted as an effect of point sources.

By studying each Galactic hemisphere separately, it was found that on
scales around $3\fdg4$, the northern hemisphere deviates significantly
from the simulations. Most importantly, the genus functional has a
very large amplitude on these scales; only 1 Monte Carlo realization
out of 5000 simulations has a larger amplitude for negative
thresholds. Interestingly, the southern hemisphere genus has a fairly
\emph{small} amplitude around the same scales. Furthermore, the spectral
parameter $\gamma$ is higher in the \emph{WMAP} data than in the
simulations on scales $\sim 3\degr$ in the northern hemisphere. These
asymmetries may be compared with those recently announced by
\citet{Eriksen:2004} and \citet{Park:2004}. Park also reports an
asymmetry in the genus amplitudes between the northern and southern
hemispheres. However, these results are found on significantly
\emph{smaller} scales than those considered here, and it is
therefore difficult to establish a direct link between our findings
and those of \citet{Park:2004}.

Many of the results presented in this paper depend on both the power
spectrum and the higher-order statistical properties of the field. For
example, the genus amplitude depends to a large extent on the power
spectrum, and the strong results found on the northern hemisphere may
therefore be influenced by both power spectrum and non-Gaussianity
issues. However, the asymmetry parameter, $\Delta g$, is a much
cleaner test of non-Gaussianity, and we found that this was extreme at
the 99\% level in the southern hemisphere in a few cases. This result
may therefore be taken as support for the results presented by
\citet{Vielva:2004}, who find non-Gaussian signatures in the southern
hemisphere using spherical wavelets. However, the \emph{WMAP} $\Delta
g$ parameter is also large at the $2\,\sigma$ level in the north around
$3\degr$ FWHM, and this is not supported by the results of
\citet{Vielva:2004}. It is therefore quite difficult to determine if
our results support the previously reported non-Gaussian effects, or
whether these findings are independent.

It is in any case clear that the currently accepted Gaussian model has
problems accounting for the statistical properties of the \emph{WMAP}
data on large and intermediate scales. In this paper we have focused
particularly on the co-added $Q+V+W$ \emph{WMAP} data, but have also shown
that the results are very stable with respect to frequencies and
different foreground correction methods. Thus, uncertainties in
foreground contributions or residuals are unlikely to resolve these
issues. While it certainly is too early to conclude that these
detections are of primordial origin, the possibility should certainly
be taken into account. The 2-year \emph{WMAP} data, and eventually the
\emph{Planck} data, will be essential in determining the nature of
these detections.

\begin{acknowledgements}
H.\ K.\ E.\ and P.\ B.\ L.\ thank the Research Council of Norway for economic
support, including a Ph.D.\ studentship for H.\ K.\ E. Some of the results
in this paper have been derived using the HEALPix software and
analysis package.  We acknowledge use of the Legacy Archive for
Microwave Background Data Analysis (LAMBDA). Support for LAMBDA is
provided by the NASA Office of Space Science.  This research used
resources of the National Energy Research Scientific Computing Center,
which is supported by the Office of Science of the US Department of
Energy under contract DE-AC03-76SF00098. This work has also
received support from the Research Council of Norway (Programme for
Supercomputing) through a grant of computing time.
\end{acknowledgements}

\end{document}